\documentclass[aip,
               jcp,
               preprint,
               amsmath,
               superscriptaddress,
               floatfix,
               nobibnotes,
               ]{revtex4-2}

\usepackage{amssymb}
\usepackage{amsmath}
\usepackage{mathtools}
\usepackage{latexsym}
\usepackage{amsfonts}
\usepackage{bm}
\usepackage[dvips]{graphicx}
\usepackage{color}
\usepackage[utf8]{inputenc}
\usepackage[flushleft]{caption}
\usepackage{subfigure}
\usepackage{rotating}
\usepackage{txfonts}
\usepackage{booktabs}
\usepackage[unicode]{hyperref}
\DeclareMathAlphabet{\mathpzc}{OT1}{pzc}{m}{it}

\def\figfoot{Multi-Layer Scheme}

\makeatletter

\makeatletter

\newcommand{\figcaption}[2]{
    \noindent {\bf Figure \ref{#1}:} #2
    \vspace{1cm}}

\begin{document}

%-------------------------------------------------------------
% Title, authors, and address
%-------------------------------------------------------------

\title{Hierarchical Wavepacket Propagation Framework via ML-MCTDH for Molecular Reaction Dynamics}

\author{Xingyu Zhang}
 \affiliation{Department of Chemistry,
              Northwestern Polytechnical University,
              West Youyi Road 127, 710072 Xi'an,
              China}

\author{Qingyong Meng}
 \email{qingyong.meng@nwpu.edu.cn}
  \affiliation{Department of Chemistry,
               Northwestern Polytechnical University,
               West Youyi Road 127, 710072 Xi'an,
               China}

\date{\today}

%-------------------------------------------------------------------------
% Abstract
%---------------------------------------------------------------------------

\begin{abstract}
{\bf Abstract}:
This work presents a computational framework for studying reaction
dynamics via wavepacket propagation, employing the multiconfiguration
time-dependent Hartree (MCTDH) method and its multilayer extension
(ML-MCTDH) as the core methodologies. The core idea centers on the
concept of modes that combine several coordinates along with their
hierarchical separations because the degrees of freedom are too numerous
to be efficiently treated as a single mode. First, the system is partitioned
into several fragments within the same layer, and these fragments are
further decomposed. Repeating this process, a hierarchical separation
of modes emerges, until modes of a manageable size are achieved. Accordingly,
the coordinates frame can be designed hierarchically. Second, the kinetic
energy operator (KEO) is derived as a sum-of-products (SOP) of single-particle
differential operators through polyspherical approach, while the potential
energy surface (PES) is expressed in a similar SOP form of single-particle
potentials (SPPs) through (1) reconstruction approaches using an existing
PES or (2) direct approaches based on a computed database. Third, the nuclear
wave function is expressed in a multi-layer expansion form, where each
term is a product of single-particle functions (SPFs) that are further
expanded by the SPFs in deeper layer. This expansion form is also
adopted by variational eigensolver for electronic wave function.
Finally, the Dirac-Frenkel variational principle leads to a set of
working equations whose solutions reproduce reaction dynamics, say
reaction probability and time-dependent expectation. In addition, the
hierarchical framework can be rearranged by the mathematical language
of tensor network (TN) or tree tensor network (TTN). In this work, we
compare the methods represented by function with those in the form of
TN or TTN. We also discuss the limitations of the present framework and
propose solutions, providing further perspectives.

\end{abstract}

\maketitle

%----------------------------------------------
% TOC
%----------------------------------------------
Table of Contents (TOC) Graphic:  \\
\begin{center}
\includegraphics[width=14cm]{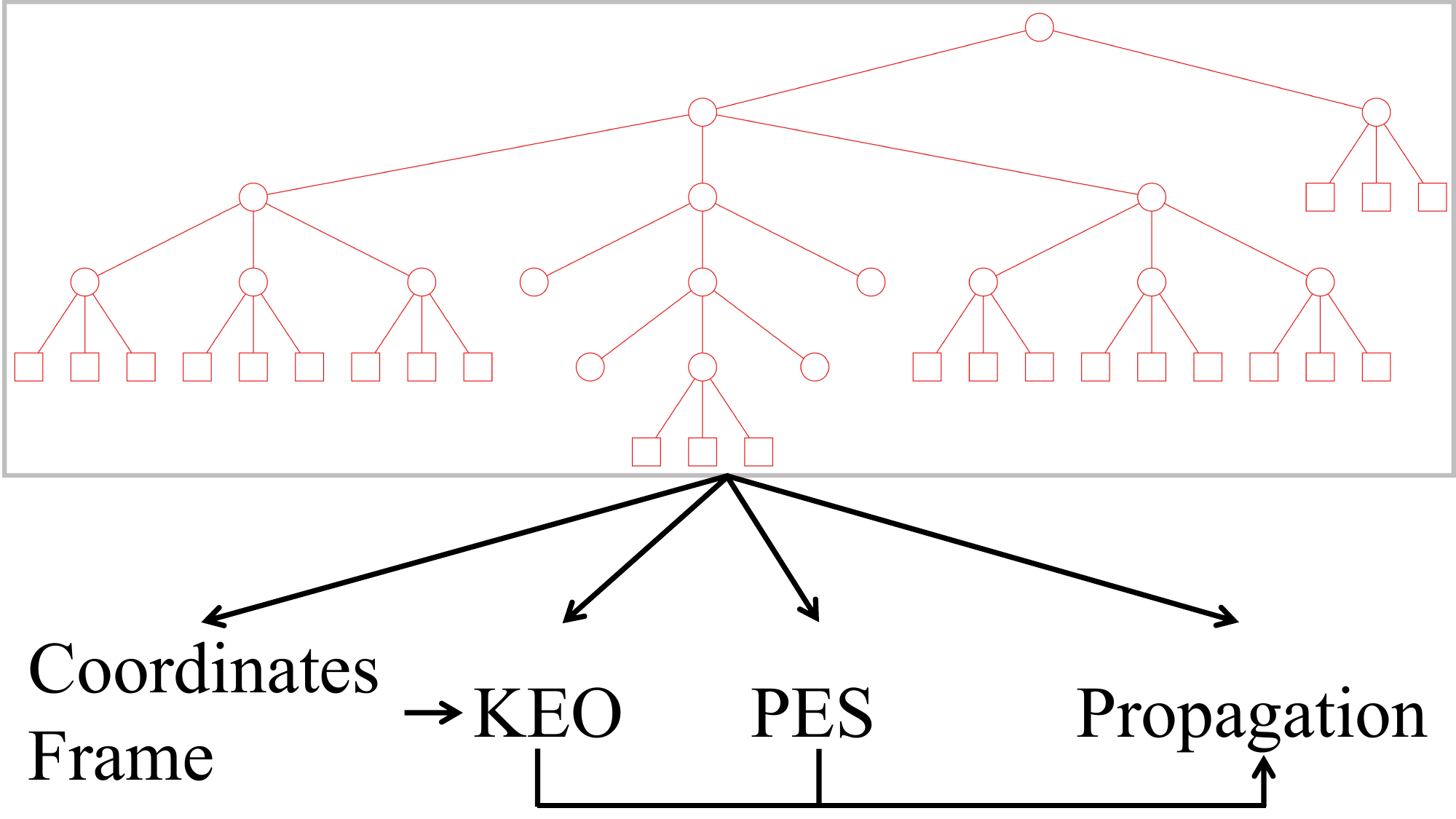}
\end{center}

\clearpage
%--------------------------------
% MAINTEXT
%--------------------------------
%%%%%%%%INTRODUCTION%%%%%%%%%%%%%%%%
\section{Introduction}

\begin{figure}
 \centering
  \includegraphics[width=14cm]{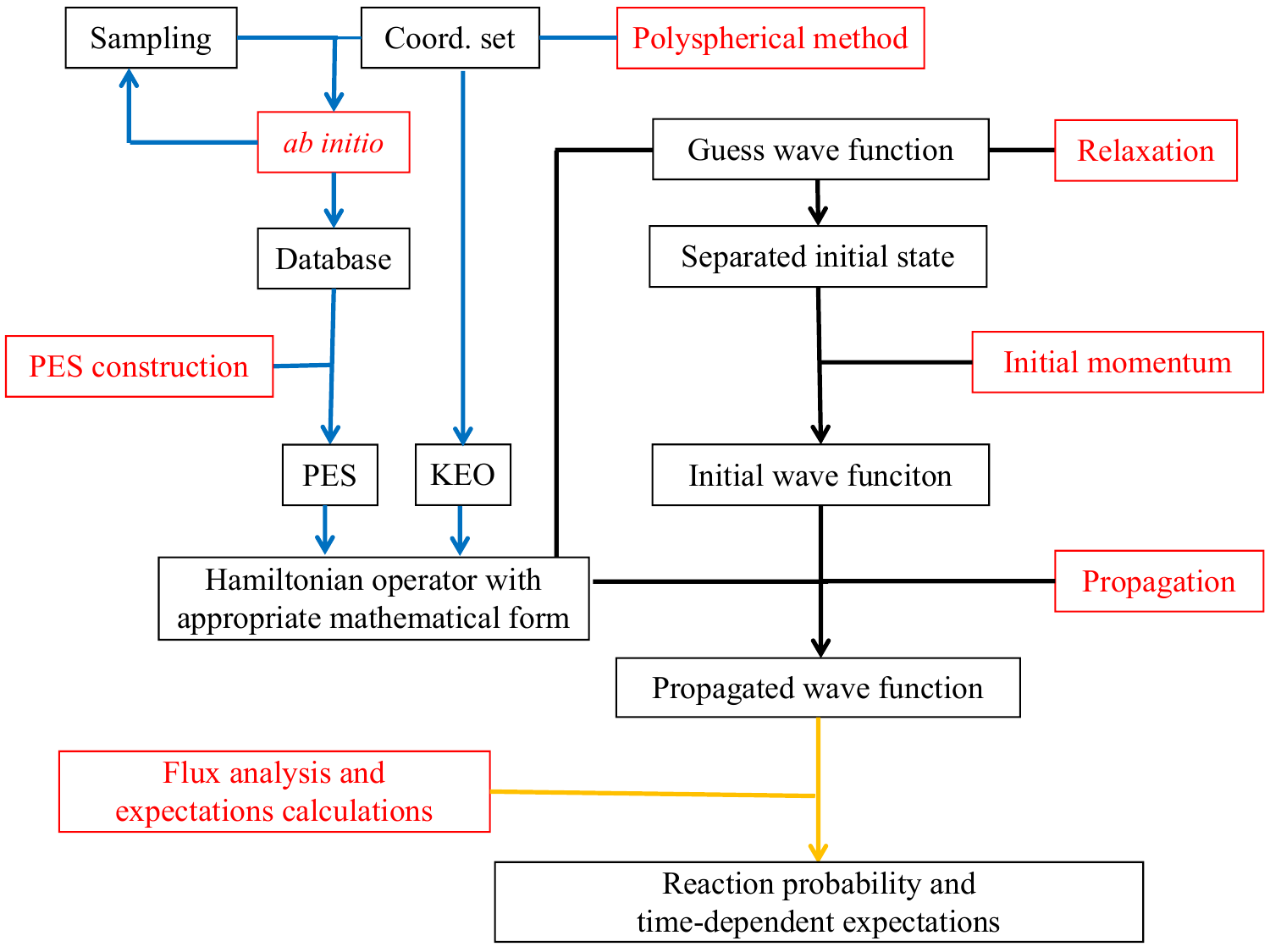}
   \caption{\figfoot}
    \label{fig:framework}
     \end{figure}

By solving the Schr{\"o}dinger equation (SE), $\mathrm{i}\partial
\Psi(t)/\partial t=\hat{H}\Psi(t)$, we can directly demonstrate chemical
dynamics from a microscopic viewpoint. Here, $\Psi(t)$ represents the
time-dependent (TD) wave function of a molecular system, and
$\hat{H}=\hat{T}+\hat{V}$ is the Hamiltonian operator, where $\hat{T}$
and $\hat{V}$ are kinetic energy operator (KEO) and potential energy
surface (PES), respectively. For a chemical reaction, we can either
examine the transformation between the eigenstates of reactant and
product, known as time-independent (TI) quantum scattering, or
investigate time-dependent (TD) wave function of the entire system,
known as wavepacket propagation. Usually, the Hamiltonian operator
$\hat{H}$ of molecular reaction system is time-independent and hence
allows the separation of time $t$ from spatial coordinates making
quantum scattering and wavepacket propagation equivalent up to a
time-energy Fourier transform. The computational methods of wavepacket
propagation, as shown by Figure \ref{fig:framework} have matured into
a sophisticated and tightly coupled framework that enables applications
to chemical dynamics. However, even elementary reactions exhibit intricate
mechanisms, posing significant theoretical and computational challenges.
This complexity necessitates the integration and coordination of the
quantum dynamics methods to achieve well performance in calculations.
A common way to achieve this goal is to express both the wave function
and the Hamiltonian operator in a hierarchical expansion form leading
to a tree-structure. The tree-structure defined by the connectivity of
the isometry single-particle objects plays a crucial role in improving
its numerical performance. This is becasue the tree-structure can
effectively enhance flexibility of the wave function and the Hamiltonian.

In this work we focus on techniques for wavepacket propagation in
molecular reaction dynamics, as shown in Figure \ref{fig:framework},
including (1) construction of $\hat{H}$ (blue arrows), (2) propagation
of an initial eigenstate $\Psi(t=0)$ (black arrows), and (3) analysis
of the propagated wave function $\Psi(t)$ (yellow arrows). A hierarchical
framework is outlined in this work to review recent developments mainly
proposed by this laboratory \cite{son24:597,mia24:532,son22:11128,men21:2702},
especially construction of the Hamiltonian operator $\hat{H}$. The core
idea centers on the concept of modes that combine several coordinates
along with their hierarchical separations, because the degrees of freedom
(DOFs) of high-dimensional systems are too numerous to be efficiently
treated as a single mode. Repeatedly breaking the mode, hierarchical
separations can be designed resulting in a multilayer representation
of the molecular system until modes of a manageable size are achieved.
In the simplest case, a $d$-dimensional system with mass-weighted
coordinates set $\mathfrak{q}=\{q^{(\kappa)}\}_{\kappa=1}^d$ has
$\hat{H}$ and $\Psi(t)$ in the summation of orthogonal products (also
called configurations) of basis, namely
\begin{align}
\hat{H}\left(q^{(1)},\cdots,q^{(d)}\right)&=
\sum_ID_I\prod_{\kappa}\hat{h}_I^{(\kappa)}\left(q^{(\kappa)}\right),
\label{eq:expansion-sop-form-00} \allowdisplaybreaks[4] \\
\Psi\left(q^{(1)},\cdots,q^{(d)},t\right)&=
\sum_IA_I(t)\prod_{\kappa}\varphi_I^{(\kappa)}\left(q^{(\kappa)},t\right),
\label{eq:expansion-sop-form-01}
\end{align}
where $\hat{h}_I^{(\kappa)}$ and $\varphi_I^{(\kappa)}$ are single-particle
operators and functions, respectively. If $q^{(\kappa)}$ is the $\kappa$-th
mode instead of coordinate, $\hat{h}_I^{(\kappa)}$ and $\varphi_I^{(\kappa)}$
can be further expanded similarly leading to multilayer sum-of-products
(SOP) form
\cite{son24:597,mia24:532,son22:11128,men21:2702,lar24:e2306881,son22:6047,mey90:73,man92:3199,mey93:141,mey98:3011,bec00:1,mey03:251,vie03:3434,wu03:14,wu04:2227,wan09:024114,wod12:124119,mey12:351,wan03:1289,man08:164116,ven11:044135,men12:134302,men13:014313,men14:124309,wan15:7951,men17:184305,man17:253001,kur18:194114,dor24:8767}.
The TD functions in the deepest layer are finally expanded either by a given
TI basis set in finite basis representation (FBR) and discrete variable
representation (DVR) \cite{bac89:469} or by a TD basis set of correlation
DVR (CDVR) \cite{har05:064106,man08:164116,man09:054109,ell22:134107,ell24:224108}
based on multidimensional time-dependent DVR (TDDVR). By CDVR, Equations
\eqref{eq:expansion-sop-form-00} and \eqref{eq:expansion-sop-form-01} are
not restrictly required. In this work, we mainly focus methods based on
the FBR or DVR required Equations \eqref{eq:expansion-sop-form-00} and \eqref{eq:expansion-sop-form-01}.

The present framework for wavepacket propagation originates from
representation of a molecular fragment by z-matrix which is well-established.
To integrate multiple z-matrices corresponding to individual fragments
into a unified representation for the whole reaction system, the
z-system was developed \cite{dix06:247,gat09:1}. For a finite set of
points in Euclidean space, this mathematical formalism \cite{dix06:247}
generalizes to a comprehensive theory of polyspherical coordinates.
Such coordinates are defined on the orbit spaces of the group of rigid
motions \cite{dix06:247}. This representation
enables the construction of hierarchical coordiante frames and
facilitates the derivation of KEO expressed as summation of
derivative product terms through polyspherical approach
\cite{lau02:8560,lau06:500,gat09:1,ndo12:034107,ndo13:204107,zha16:204302}. Following
Equation \eqref{eq:expansion-sop-form-00}, significant developments
have been achieved to construct the PES in a sum-of-products (SOP)
form either by an existing PES in general form or by a set of energy
values at sampled geometries. Employing Equations \eqref{eq:expansion-sop-form-00}
and \eqref{eq:expansion-sop-form-01}, various propagation approaches
have been developed over the past decades. Depending on the number of
layers in the expansional ansatz, these approaches are convention (or
known as standard) method, time-dependent Hartree (TDH) product,
multiconfiguration TDH (called MCTDH) and multilayer MCTDH (called
ML-MCTDH). Due to flexibility of the wave function ansatz, both MCTDH
and ML-MCTDH methods have greatly influenced the field of quantum
molecular dynamics and continue to do so. In this work, we will mainly
discuss the hierarchical wavepacket propagation for molecular reaction
dynamics using the MCTDH language which employs the Schr{\"o}dinger
picture with a grid-based coordinate representation. This is a natural
choice for chemists who want to directly demonstrate how a chemical reaction
occurs. Moreover, grid-based representations of wave function and PES
yield tensor-structured expressions, analogous to tensor network (TN)
or tree tensor network (TTN) in quantum many-body systems \cite{shi06:022320},
such as matrix product state (MPS) and matrix product operator (MPO).
By such tensor-structured representation, the time-dependent variational
principle indicates the time-dependent density matrix renormalization
group (TD-DMRG) method
\cite{whi04:076401,dal04:P04005,gui04:040502,adr05:020404,jua06:305,has06:195304,ron17:5560,iou18:134115,shi18:174102,bai19:3481,dan23:044101,bai24:3151}
or a variant of MCTDH, called MPS-MCTDH \cite{kur18:194114}.
This is not surprising because chemical dynamics is fundamentally a
many-body problem and the current ML-MCTDH-based framework employs the
single-particle approximation (SPA).

Furthermore, in this work we compare the wavepacket propagation methods
with the electronic-structure methods because the antisymmetric wave
function in the SOP form is widely adopted by variational approaches
for electronic structure \cite{sza96:book,sza15:1342}. These methods
are Hartree-Fock (HF), full configuration interaction (CI), multiconfigurational
self-consistent field (MCSCF), and so forth. The one-dimensional basis
function in the deepest layer is called molecular orbital (MO) which is
further expanded by a given set of basis functions. Since these
electronic-structure methods also employ the SPA, substituting the
antisymmetry MPS as ansatz into the variational principle, one can
derive working equations for quantum chemistry (QC) time-independent
density matrix renormalization group (TI-DMRG or called QC-DMRG)
\cite{whi92:2863,whi93:10345,cha12:907,wou14:272,sza15:1342}. 
The TI-DMRG method is a special case of time-independent TN (TI-TN)
optimizing a one-dimensional (1D) MPS but not a time-independent TTN
(TI-TTN). In the following, we collectively refer to TD-DMRG and TI-DMRG
as DMRG. Initially, ML-MCTDH and DMRG focused on very different applications.
The former is based on quantum molecular dynamics \cite{wan03:1289,man08:164116,ven11:044135}
whereas the latter is based on the quantum many-body dynamics \cite{whi92:2863,whi93:10345}.
Approximately two decades after their inceptions, both ML-MCTDH and
DMRG were extended to incorporate the capability of solving each other's
problems. The different individual developments of ML-MCTDH and DMRG
led to very different mathematical expressions. Recently, Larsson
\cite{lar24:e2306881} mediated the communicative barriers between the
ML-MCTDH and TD-DMRG formulations by deriving working equations through
counterpart's language.

In this work, to demonstrate the present hierarchical framework for
wavepacket propagation, vibrational eigenstates and sticking probability
of CO/Cu(100) \cite{men21:2702}, dissociative chemisorption probability
of H$_2$O/Cu(111) \cite{son22:6047}, and reaction probability of OH +
HO$_2$ $\to$ O$_2$ + H$_2$O (denoted by HO$_x$) \cite{son24:597} are
taken as examples. Vibrational eigenstates and the probabilities of
chemical reactions or physical processes are central problems in
chemical dynamics. We refer the reader to the Supporting Information
for implementations and numerical details of these examples. The paper
is organized as follows; in Sections \ref{sec:coord-keo} and \ref{sec:pes-cpd},
we describe construction approaches of KEO and PES, respectively.
Section \ref{sec:working} describes function expansions for wavepacket
propagation and compares these with those for electronic structure.
Section \ref{sec:disscuss} presents extensions of the present theory.
Section \ref{sec:conclud} concludes with a summary.

%%%%%%%%%%%%%%%%%%%%%%%%%%%%%%%%%%%%%%%%
\section{Coordinates Set and Kinetic Energy Operator\label{sec:coord-keo}}

To easily derive the KEO, Cartesian coordinates set is the simplest
choice to represent the molecular system because all of coordinates
are orthogonal. For $N$-atomic system, multiplying square root of
associated mass $\{m_{\alpha}\}_{\alpha=1}^N$ mass-weighted Cartesian
coordinates set $\mathfrak{x}_N=\{x_{\alpha},y_{\alpha},z_{\alpha}\}_{\alpha=1}^N$
indicates the KEO
\begin{equation}
\hat{T}=-\frac{1}{2}\sum_{\alpha=1}^N
\left(\frac{\partial^2}{\partial x_{\alpha}^2}+
\frac{\partial^2}{\partial y_{\alpha}^2}+
\frac{\partial^2}{\partial z_{\alpha}^2}\right),
\label{eq:keo-cartesian-f-dofs-99}
\end{equation}
which is automatically given in the SOP form.
Unless otherwise specified, in this section mass-weighted coordinates
are always used. However, the Cartesian coordinates set is rarely
employed in quantum dynamics because chemical processes with breaking
and forming of the chemical bonds cannot be intuitively described by
such coordinates set. Moreover, Cartesian coordinates always introduce
artificial correlations slowing down computational convergence. To
overcome this question, generalized coordinates are designed as linear
or non-linear functions of the Cartesian coordinates. A set of linear
functions of $\mathfrak{q}_N$ is rectilinear coordinates set. A typical
example is vibrational normal mode of a bound system. A set of non-linear
functions of $\mathfrak{q}_N$ is curvilinear coordinates set which is
able to describe rotations and motions with large amplitude. Both
rectilinear and curvilinear coordinates might be involved in describing
chemistry dynamics. Therefore, the former is well-suited for quantum
dynamics calculations of spectra, while the latter is appropriate for
molecular reaction dynamics.

Over the past several decades, the polyspherical approach has been
developed to design an appropriate set of curvilinear coordinates and
to derive the KEO \cite{gat09:1}. Its core idea involves partitioning
the system into multiple subsystems of sufficient simplicity to
facilitate coordinate definition, while employing Euler angles to
describe the relative orientations between subsystems and the system
itself. Moreover, such separation allows the reaction coordinate to be
naturally defined as a subset of large-amplitude curvilinear coordinates.
Figure \ref{fig:hier-coord} illustrates the hierarchical scheme for
designing coordinates through recursive fragmentation of the system
into fragments ({\it i.e.}, subsystems), corresponding to the decomposition
of modes into deeper layers. Comparing with the previously reported
polyspherical approach \cite{gat09:1,ndo12:034107,ndo13:204107}, the
present hierarchical scheme is given in multilayer fashion instead of
one-layer. As shown in Figure \ref{fig:hier-coord}, the translational
and rotational motions of the entire system are described in the
space-fixed (SF) frame or called the laboratory-fixed (LF) frame,
whose axes are fixed at the laboratory. Similarly fixing axes at the
system itself, we obtain the E$_2$ frame in which the internal motions
are described. Second, to describe internal motions of individual
fragments, body-fixed (BF) frames are established by fixing their
axes to specific fragments. The relative orientations between each BF
frame and its parent frame in the last layer are specified through
three Euler angles $\{\alpha^{(\mathrm{BF})},\beta^{(\mathrm{BF})},\gamma^{(\mathrm{BF})}\}$
that form a relative frame. Repeatedly introducing BF frames with respect
to those in the last layer, a hierarchical frame is designed. The deepest
layer BF frames describe the smallest inactive fragments during reaction
using minimal coordinates, typically represented by Jacobi or Radau
vectors \cite{zha16:204302}. Figure \ref{fig:ml-expansion-examples}
illustrates the applications of the hierarchical coordinate frame
to the present benchmarks, CO/Cu(100) \cite{men21:2702}, H$_2$O/Cu(111)
\cite{son22:6047}, and HO$_x$ \cite{son24:597}. These coordinates
frames are all designed in one layer.

Having mass-weighted coordinates $\mathfrak{q}=\{q^{(\kappa)}\}_{\kappa=1}^d$
of a $d$-dimensional system, the KEO can be derived through the metric
tensor $g^{\kappa\rho}$ and Christoffel symbol of the second kind
$\Gamma_{\kappa\rho}^{\upsilon}$, \cite{lau02:8560,lau06:500,gat09:1,ndo12:034107,ndo13:204107}
\begin{equation}
\hat{T}=-\frac{1}{2}\sum_{\kappa,\rho=1}^dg^{\kappa\rho}
\left(\frac{\partial^2}{\partial q^{(\kappa)}\partial q^{(\rho)}}
-\sum_{\upsilon=1}^d\Gamma_{\kappa\rho}^{\upsilon}
\frac{\partial}{\partial q^{(\upsilon)}}\right)
=-\frac{1}{2}\sum_{\kappa,\rho=1}^dg^{\kappa\rho}\Big(\partial_{\kappa}\partial_{\rho}-
\sum_{\upsilon=1}^d\Gamma_{\kappa\rho}^{\upsilon}\partial_{\upsilon}\Big).
\label{eq:polysph-keo-003}
\end{equation}
Both $g^{\kappa\rho}$ and $\Gamma_{\kappa\rho}^{\upsilon}$ are determined
by the coordinate transformations between $\{q^{(\kappa)}\}_{\kappa=1}^d$
and standard Cartesian coordinates. Several numerical procedures, say
the TNUM (means numerical $\hat{T}$) procedure \cite{lau02:8560,lau06:500},
were developed to obtain $g^{\kappa\rho}$ and $\Gamma_{\kappa\rho}^{\upsilon}$
and their derivatives and to finally compute numerical KEOs. Moreover,
employing the polyspherical approach, an automatic procedure
\cite{ndo12:034107,ndo13:204107} was developed to provide explicit
analytical expression of the KEO which can be adopted by MCTDH and
ML-MCTDH. In a word, the starting point of these numerical or procedures
is the relation between the mass-weighted Cartesian coordinates in the
BF frame and the curvilinear coordinates. For the $(ij\cdots pk)$-th
BF frame (denoted as BF$_{ij\cdots pk}$ in Figure \ref{fig:hier-coord})
associated with an $N_{ij\cdots pk}$-atomic fragment in the deepest
layer, the corresponding KEO $\hat{T}^{(\mathrm{BF}_{ij\cdots pk})}$
can be expressed in terms of $3N_{ij\cdots pk}-6$ momentum operators
$\{\hat{P}_m\}_{m=1}^{3N_{ij\cdots pk}-6}$ and three angular momenta
$\{\hat{J}_{\iota}\}_{\iota=1}^3$. This operator decomposes into
vibrational ({\it i.e.} $\hat{T}_{\mathrm{vib}}^{(\mathrm{BF}_{ij\cdots pk})}$),
rotational ({\it i.e.} $\hat{T}_{\mathrm{rot}}^{(\mathrm{BF}_{ij\cdots pk})}$),
and Coriolis coupling ({\it i.e.} $\hat{T}_{\mathrm{cor}}^{(\mathrm{BF}_{ij\cdots pk})}$)
components, \cite{lau02:8560,lau06:500,gat09:1,ndo12:034107,ndo13:204107}
\begin{align}
2\hat{T}^{(\mathrm{BF}_{ij\cdots pk})}&
=2\hat{T}_{\mathrm{vib}}^{(\mathrm{BF}_{ij\cdots pk})}
+2\hat{T}_{\mathrm{cor}}^{(\mathrm{BF}_{ij\cdots pk})}
+2\hat{T}_{\mathrm{rot}}^{(\mathrm{BF}_{ij\cdots pk})}
\allowdisplaybreaks[4] \nonumber \\
&=\sum_{m,m'=1}^{3N_{ij\cdots pk}-6}\hat{P}^{\dagger}_m
\Sigma_{mm'}\hat{P}_{m'}+
\sum_{m=1}^{3N_{ij\cdots pk}-6}\sum_{\iota=1}^3\Big(
\hat{P}^{\dagger}_m\sigma_{m\iota}\hat{J}_{\iota}+
\hat{J}_{\iota}^{\dagger}\sigma_{\iota m}^{\dagger}\hat{P}_m\Big)
+\sum_{\iota,\varkappa=1}^3\Big(
\hat{J}_{\iota}^{\dagger}\Gamma_{\iota\varkappa}\hat{J}_{\varkappa}+
\hat{J}_{\varkappa}^{\dagger}\Gamma_{\varkappa\iota}^{\dagger}\hat{J}_{\iota}\Big).
\label{eq:polysph-keo-022}
\end{align}
The parameter tensors include (1) the $(3N_{ij\cdots pk}-6)\times
(3N_{ij\cdots pk}-6)$ matrix $\boldsymbol{\Sigma}$, (2) the
$(3N_{ij\cdots pk}-6)\times3$ matrix $\boldsymbol{\sigma}$, and (3)
the $3\times3$ matrix $\boldsymbol{\Gamma}$. It is often impossible
to completely separate these three terms and the splitting itself is
artificial \cite{gat09:1}. This is because the Coriolis coupling and
the rotational part depend on the BF$_{ij\cdots pk}$ frame. Following
the derivation of $\hat{T}^{(\mathrm{BF}_{ij\cdots pk})}$, the KEO in
the BF$_{ij\cdots p}$ frame can be written as a summation of
$\hat{T}^{(\mathrm{BF}_{ij\cdots pk})}$ and the KEOs from relative
frames \cite{gat09:1}. Equation \eqref{eq:polysph-keo-022} in a
summation of products of coordinate-associated operators, naturally
leads to a multi-layer expansion form for the final KEO expression.

Main advantages of the hierarchical coordiante frame are (1) its clear
descriptions of the chemical and physical insights in a reaction, and
(2) its capability to automatically \cite{lau02:8560,lau06:500,gat09:1,ndo12:034107,ndo13:204107}
derive the KEO in the summation of products of mono-mode operators
through the polyspherical approach. First, the reaction dynamics
involves large-amplitude motions which are described by curvilinear
coordinates. By the hierarchical frame, one can simply define the
curvilinear coordinates for large-amplitude motions and hence define
the reaction through chemical inspirations. It is generally accepted
that a given reaction coordinate maps to a specific reaction channel
or mechanism. Consequently, carefully designed reaction coordinates
based on chemical inspirations enable both (1) more efficient sampling
of geometries in configurational space for PES construction (see Section
\ref{sec:pes-cpd}) and (2) deeper analysis of the corresponding reaction
dynamics. This represents a crucial feature when dealing with large
elementary-reactions with complex dynamcis mechanisms. For instance,
chemical inspirations indicate that the HO$_x$ system has four channels
and thus three reactions starting from OH + HO$_2$,
\begin{equation}
\mathrm{OH}+\mathrm{HO}_2\to\mathrm{O}_2+\mathrm{H}_2\mathrm{O},\quad
\mathrm{OH}+\mathrm{HO}_2\to\mathrm{H}_2+\mathrm{O}_3,\quad
\mathrm{OH}+\mathrm{HO}_2\to\mathrm{O}+\mathrm{H}_2\mathrm{O}_2.
\label{eq:o2-wat}
\end{equation}
Obviously, coordinates defined in Figure \ref{fig:ml-expansion-examples}(c)
are only suitable to describe dynamics of the
$\mathrm{OH}+\mathrm{HO}_2\to\mathrm{O}_2+\mathrm{H}_2\mathrm{O}$ reaction.
In addition, description with individual fragment allows us to use direct
products of one-dimenional basis avoiding the singularity problem. This
also leads to a reduced coupling between the coordinates. Second, for
one-layer separation scheme, Lauvergnat and co-workers \cite{lau02:8560,lau06:500}
as well as Ndong and co-workers \cite{ndo12:034107,ndo13:204107}
have developed procedures to derive the KEO in numerical or analytical
form. For hierarchical frame, it is also possible to develop similar
technique. Currently, however, a single-layer scheme is sufficient to
describe typical elementary reactions. The hierarchical frame incorporates
an interface for future consideration of substituent motions based on
the one-layer scheme.

One notable disadvantage of the hierarchical coordiante frame is the
redundancy in Euler angles. Specifically, when two
relative frames share a common axis (typically denoted by the $z$-axis),
the two Euler angles around this axis depend on each other forming a
pair of redundancy azimuth angles, called $\{\alpha_i\}_{i=1}^2$. To
address this issue, one of the authors \cite{men13:164709} proposed a
solution by applying a Fourier transform to $\{\alpha_i\}_{i=1}^2$,
converting them into their discrete momentum counterparts. Consequently,
the corresponding KEO terms $\{\partial^2/\partial\alpha_i^2\}_{i=1}^2$
are replaced by the square of the angle quantum numbers $\{k_i^2\}_{i=1}^2$.
Moreover, the action of $\hat{V}$ on $\Psi$ is accordingly written as
an expansion form in the angular-momentum representation
\cite{men13:164709},
\begin{equation}
V\Big(\mathfrak{q}',\alpha_1,\alpha_2\Big)
=\sum_{\Omega^{(1)}=-k_m^{(1)}}^{k_m^{(1)}}
\sum_{\Omega^{(2)}=-k_m^{(2)}}^{k_m^{(2)}}
\tilde{V}_{\Omega^{(1)},\Omega^{(2)}}\Big(\mathfrak{q}'\Big)
\exp\Big(i\Omega^{(1)}\alpha_1\Big)\exp\Big(i\Omega^{(2)}\alpha_2\Big),
\label{eq:fourier-transform-000}
\end{equation}
where $\mathfrak{q}'=\mathfrak{q}\setminus\{\alpha_i\}_{i=1}^2$ and
$\tilde{V}_{\Omega^{(1)},\Omega^{(2)}}(\mathfrak{q}')$ satisfies
Fourier transform
\begin{equation}
\tilde{V}_{\Omega^{(1)},\Omega^{(2)}}\Big(\mathfrak{q}'\Big)
\propto\iint_0^{2\pi}\exp\Big(-i\Omega^{(1)}\alpha_1\Big)
\exp\Big(-i\Omega^{(2)}\alpha_2\Big)
V\Big(\mathfrak{q}',\alpha_1,\alpha_2\Big)
\mathrm{d}\alpha_1\mathrm{d}\alpha_2.
\label{eq:fourier-transform-001}
\end{equation}
Another disadvantage arises from the non-inertial nature of the
hierarchical coordinate frame. This issue is not new. The ancient
geocentric theory describes planetary motions using a non-inertial
frame with the epicycle-on-deferent technique. The epicycle was
defined to move in longitude, while the planet moves on the epicycle
at a specific velocity \cite{han60:150}. Similarly, the present
hierarchical frame represents arbitrary periodic motions through the
superposition of a set of circular motions, which introduces numerical
complexity and reduces the intuitiveness of the resulting KEO. A
potential solution to mitigate this problem is to adopt an occupancy
representation rather than a coordinate representation implemented in
a first-quantization (FQ) framework. By the second-quantization (SQ)
formalism (shown in Section \ref{sec:disscuss} later), the atomic motions
are represented by excitations of virtual particles (or fields). The
kinetic energy term is fundamentally a representation of the single-particle
KEO expressed in a chosen basis. In other words, it is constructed from
matrix elements of the single-particle KEO in the selected basis. This
is helpful to simplify the KEO term.

%-------------------PES----------------
\section{Decomposition of Potential Energy Surface\label{sec:pes-cpd}}

Since the KEO has already been given in the SOP form, as given by Equation
\eqref{eq:polysph-keo-022}, now the remaining task to build the Hamiltonian
operator $\hat{H}$ in the form of Equation \eqref{eq:expansion-sop-form-00}
is to construct the PES in sum-of-products (SOP) or canonical-polyadic (CP)
decomposition (CPD) of single-particle potentials (SPPs),
\begin{equation}
V^{(\mathrm{SOP})}\big(\mathfrak{q}\big)=\sum_{i_1=1}^{m_1}\cdots
\sum_{i_d=1}^{m_d}C_{i_1\cdots i_d}\prod_{\kappa=1}^d
v_{i_\kappa}^{(\kappa)}\big(q^{(\kappa)}\big),\quad
V^{(\mathrm{CPD})}\big(\mathfrak{q}\big)=\sum_{r=1}^RC_r\prod_{\kappa=1}^d
v_r^{(\kappa)}\big(q^{(\kappa)}\big),
\label{eq:cpd-sop-form-4-pes}
\end{equation}
where $C_{i_1\cdots i_d}$ and $C_r$ are expansion coefficients or called
coefficient tensor, $R$ the CPD rank, and $v_i^{(\kappa)}(q^{(\kappa)})$
the $i$-th SPP for the $\kappa$-th coordinate. If $q^{(\kappa)}$ means
the $\kappa$-th mode, $v_i^{(\kappa)}(q^{(\kappa)})$ can be further
expanded similarly leading to multi-layer expansion form. For the
simplest one-layer expansion form of Equation \eqref{eq:cpd-sop-form-4-pes},
two primary ways can be employed. One way is direct construction from
an energy database, such as reproducing kernel Hilbert space interpolation (RKHSI) \cite{hol97:7223}
and revised Gaussian process regression (GPR) proposed by this laboratory
\cite{son24:597,son22:11128}. Another way is reconstruction using an
existing PES, say potential re-fitting (POTFIT) method \cite{jae95:5605,jae96:7974,jae98:3772}.
We collect these methods in Table \ref{tab:methods-sop-cpd}. On the
other hand, if MPS ansatz is applied to represent the wave function,
matrix product operator (MPO) \cite{ver04:207204,pir10:025012} as tensor
network operator should be adopted to represent potential tensor on grids.
The conventional MPO is designed to handle a SOP of one-body operators
such as position operators and second quantization operators. An efficient
algorithm for compressing a given PES into a grid-based MPO was recently
developed \cite{hin24:3839} to work with MPS-MCTDH, where the MPO is
constructed from the grid-based numerical integral of the sum of arbitrary
many-body coupled operators. It has capability to provide numerical
integrals of arbitrary potential operator representations. The potential
tensors with the SOP or CPD form as given in Equation \eqref{eq:cpd-sop-form-4-pes}
represent two of tensor decomposition approaches
\cite{car70:283,tuc66:279,dol21:A2190,hac09:706,lub13:470,kol09:455,ose10:70,avi15:044106,bar15:174107,gor19:59,str24:54,gor08:404} 
for the approximation of functions. In Table \ref{tab:tensor-decomp-pes}
we briefly summarize tensor decomposition approaches for the PES construction.

First of all, the direct approaches build the potential function in the
SOP or CPD form through a discrete set of energy data. Originating from
generalized linear regression (GLR), the neural network
(NN) approach has gained widespread adoption in building the PES owing
to its universal approximation capability for continuous functions. It
is this remarkable property which motivated Manzhos and Carrington
\cite{man06:194105} and Koch and Zhang \cite{koc14:021101} to propose
a single-layer NN architecture in the SOP/CPD form, subsequently termed
SOP-NN,
\begin{equation}
V^{(\mathrm{SOP-NN})}(\mathfrak{q})=b^{(2)}+\sum_{k=1}^{F^{(1)}}
c_k^{(2)}\prod_{\kappa=1}^df\Big(b_{k\kappa}^{(1)}+
w_{k\kappa}^{(1)}q^{(\kappa)}\Big),
\label{eq:nn-sop-form-000}
\end{equation}
where $b^{(2)}$ and $c_k^{(2)}$ are parameters, while $b_{k\kappa}^{(1)}$
and $w_{k\kappa}^{(1)}$ are biases and weights, respectively. The
$V^{(\mathrm{SOP-NN})}$ formulation in Equation \eqref{eq:nn-sop-form-000}
can be extended to a multi-layer architecture by sequentially inserting
the output of one neuron into the activation function of the subsequent
layer. However, this nested structure inherently loses the CPD characteristic.
This limitation can be resolved through a multi-layer SOP-NN approach,
called ML-SOP-NN, where $q^{(\kappa)}$ in Equation
\eqref{eq:nn-sop-form-000} represents the $\kappa$-th mode rather than
a coordinate. The feasibility of this approach stems from the fact that
the utilization of modes and the mode combination \cite{bec00:1}
preserve the essential characteristics required for PES construction.
By iteratively incorporating neurons into activation functions following
a hierarchical mode structure, one can derive the PES in a multi-layer
expansion form, denoted by $V^{(\mathrm{ML-SOP-NN})}$.

It has been proved that the kernel-model regression (KMR), such as GPR,
is equivalent to single-layer NN with infinite neurons \cite{son22:1983}.
Inspiring with SOP-NN and RKHSI \cite{hol97:7223}, this laboratory
proposed a CPD-type GPR function, called CPD-GPR \cite{son22:11128},
and its revisions with mode-combination (mc) scheme \cite{bec00:1},
called CPD-mc-GPR \cite{son24:597},
\begin{equation}
V^{(\mathrm{CPD-GPR})}\big(\mathfrak{q}\big)=
\sum_{r=1}^n\xi_rK\big(\mathfrak{q},\overline{\mathfrak{q}}_r\big)
=\sum_{r=1}^n\xi_r\prod_{\kappa=1}^d
K^{(\kappa)}\big(q^{(\kappa)},\overline{q}^{(\kappa)}_r\big),\quad
K\big(\mathfrak{q},\overline{\mathfrak{q}}_r\big)=\prod_{\kappa=1}^d
K^{(\kappa)}\big(q^{(\kappa)},\overline{q}^{(\kappa)}_r\big),
\label{eq:cpd-gpr-4cpd}
\end{equation}
where $K^{(\kappa)}(\cdot,\cdot)$ is the kernel function for the $\kappa$-th
coordinate, $\xi_r$ the $r$-th optimized coefficient,
$\{\overline{\mathfrak{q}}_r\}_{r=1}^n=\{\overline{q}^{(1)}_r,
\cdots,\overline{q}^{(d)}_r\}_{r=1}^n$ geometries of $n$ traning data.
Since the mode combination scheme \cite{bec00:1} does not affect the
construction of PES, we no longer need to distinguish CPD-GPR and
CPD-mc-GPR. Equation \eqref{eq:cpd-gpr-4cpd} indicates that
$V^{(\mathrm{CPD-GPR})}$ has a total of $n$ expansion terms (rank)
equal to number of tranining data \cite{son22:11128,son24:597}. This
value is often larger than $10^4$ and leads to low-performance in
subsequent calculations \cite{son22:11128,son24:597}. To reduce rank,
method to select decisive data was developed by this laboratory through
support vector regression (SVR) in the CPD form, called CPD-SVR
\cite{son22:1983,mia24:2410.23529}. The resulting PES is similar to
Equation \eqref{eq:cpd-gpr-4cpd} but $n$ is replaced by number of
support vectors $n_{\mathrm{sv}}<n$, \cite{son22:1983,mia24:2410.23529}
\begin{equation}
V^{(\mathrm{CPD-SVR})}\big(\mathfrak{q}\big)=
\sum_{r=1}^{n_{\mathrm{sv}}}\xi_r
K\big(\mathfrak{q},\overline{\mathfrak{q}}_r\big)+b
=\sum_{r=1}^{n_{\mathrm{sv}}}\xi_r\prod_{\kappa=1}^d
K^{(\kappa)}\big(q^{(\kappa)},\overline{q}^{(\kappa)}_r\big)+b,\;
K\big(\mathfrak{q},\overline{\mathfrak{q}}_r\big)=\prod_{\kappa=1}^d
K^{(\kappa)}\big(q^{(\kappa)},\overline{q}^{(\kappa)}_r\big),
\label{eq:cpd-gpr-4cpd-000}
\end{equation}
where intercept $b$ is required by selection algorithm \cite{son22:1983,mia24:2410.23529}.
To easily select support vectors, an existing PES is introduced as
initial decision surface to warm-start (ws) the selection process.
This is the CPD-ws-SVR method \cite{mia24:2410.23529}. Concerning on
the resulting PES expression, we will no longer distinguish CPD-SVR
and CPD-ws-SVR, as well as CPD-mc-SVR and CPD-ws-mc-SVR. According to
hierarchical coordinate frame, one can repeatedly substitute CPD-GPR/SVR
function into $V^{(\mathrm{CPD-GPR/SVR})}$ to obtain multi-layer version
of CPD-GPR/SVR, called ML-CPD-GPR/SVR. These methods for the system with
$l$ layers predict the ranks of $l\cdot n$ and $l\cdot n_{\mathrm{sv}}$
which are much larger than $n$ and $n_{\mathrm{sv}}$. Nevertheless, the
ML-CPD-GPR/SVR function might be more in line with the multi-layer wave
function and easier to simplify the working equations (see Section
\ref{sec:working} later).

Next, we turn to the reconstruction approaches, as collected in Table
\ref{tab:methods-sop-cpd}. One of typical methods is POTFIT \cite{jae95:5605,jae96:7974,jae98:3772}
which transfers the PES with less than six dimensionalities into the SOP
form. For higher-dimension (larger than six), the multi-grid POTFIT (MGPF)
\cite{pel13:014108}, multi-layer POTFIT (MLPF) \cite{ott14:014106}, and
Monte Carlo POTFIT (MCPF) \cite{sch17:064105} methods can be adopted to
build the SOP form, while the Monte Carlo CPD re-fitting (MCCPD) method
\cite{sch20:024108,men21:2702} is useful to build high-dimensional CPD
functions. For instance, by MCCPD, the PESs of CO/Cu(100) \cite{men21:2702}
and H$_2$O/Cu(111) \cite{son22:11128} were re-fitted in this laboratory
showing its power. Gatti and co-workers \cite{shi23:194102,shi25:1896}
performed 75D MCCPD calculations for the PES of a hydrogen atom scattering
from graphene. Moreover, Pel{\'a}ez and co-workers
\cite{pan20:234110,nad23:114109} proposed a novel FBR formulated
in the SOP/CPD form, denoted by SOP/CP-FBR. These methods enable the
interpolation of a PES from coarse grids to fine grids, provided that
the SPPs are represented by analytical functions and expanded through
a given set of polynomial series, such as Chebyshev polynomial series
$\{T_{\mu}(\cdot)\}_{\mu=1}^{s}$, \cite{pan20:234110,nad23:114109}
\begin{align}
V^{(\mathrm{SOP-FBR})}\big(\mathfrak{q}\big)
&=\sum_{j_1=1}^{m_1}\cdots\sum_{j_d=1}^{m_d}
C_{j_1,\cdots,j_d}\prod_{\kappa=1}^d\left(\sum_{\mu=1}^{s_{\kappa}}
c_{\mu,j_{\kappa}}^{(\kappa)}T_{\mu}\big(q^{(\kappa)}\big)\right),
\label{eq:sop-fbr-000}
\allowdisplaybreaks[4]  \\
V^{(\mathrm{CP-FBR})}\big(\mathfrak{q}\big)&=\sum_{r=1}^RC_r
\prod_{\kappa=1}^d\left(\sum_{\mu=1}^{s_{\kappa}}
c_{\mu,r}^{(\kappa)}T_{\mu}\big(q^{(\kappa)}\big)\right).
\label{eq:sop-fbr-001}
\end{align}
Here, $c_{\mu,j_{\kappa}}^{(\kappa)}$ and $c_{\mu,r}^{(\kappa)}$ are
expansion coefficients. Due to the sparse character of the coefficient
tensors (called Tucker core tensors), the SOP/CP-FBR methods
\cite{pan20:234110,nad23:114109} save computational cost of the
subsequent propagations. It is not surprising that, if kernel
functions of CPD-GPR/SVR are expanded by Chebyshev basis functions
with infinite order $s_{\kappa}\to\infty$, then CPD-GPR/SVR become
CP-FBR.

Next, we must turn to the tensor decomposition approaches (see Table
\ref{tab:tensor-decomp-pes}) for building the potential tensor in the
SOP or CPD form. In propagating the wave function, its discretization
requires tensor representation of the PES. In this context, the PES is
represented by the tensor with $d$-way array. For instance, the potential
functions shown by Equation \eqref{eq:cpd-sop-form-4-pes} can be represented
by their values on grids $I=(i_1,\cdots,i_d)$,
\begin{equation}
V_I^{(\mathrm{SOP})}=V^{(\mathrm{SOP})}_{i_1\cdots i_d}=
V^{(\mathrm{SOP})}(q_{i_1}^{(1)}\cdots,q_{i_{\kappa}}^{(\kappa)},\cdots,q_{i_d}^{(d)}),\quad
V_I^{(\mathrm{CPD})}=V^{(\mathrm{CPD})}_{i_1\cdots i_d}=
V^{(\mathrm{CPD})}(q_{i_1}^{(1)}\cdots,q_{i_{\kappa}}^{(\kappa)},\cdots,q_{i_d}^{(d)}),
\label{eq:cpd-sop-form-999}
\end{equation}
where $q_{i_{\kappa}}^{(\kappa)}$ means values of the $i_{\kappa}$-th
grid for the $\kappa$-th DOF. The tensor decompositions $V_I^{(\mathrm{SOP})}$
and $V_I^{(\mathrm{CPD})}$ are called Tucker and CP decompositions,
respectively. The Tucker decomposition is a higher-order form of
principal component analysis while the CP decomposition is a sum of
rank-one tensors \cite{kol09:455}. Other tensor decomposition methods
include tensor train (TT) decomposition, hierarchical Tucker (HT)
decomposition, cross-approximation method, and so forth. In Table
\ref{tab:comp-tensor-function-99}, we compare the methods given in
Table \ref{tab:methods-sop-cpd} with the tensor decomposition approaches
(see Table \ref{tab:tensor-decomp-pes}) for fitting the potential
tensor in the sum of products of tensors. Baranova
and Oseledets \cite{bar15:174107} applied the TT cross approximation
procedure to the PES construction, in conjucntion with an affine
transformation of Cartesian coordinates into the active subspaces (AS)
where the PES has the most variability. This TT+AS approach \cite{bar15:174107}
belongs to reconstruction approach which requires an existing PES
or couples to a quantum chemistry code. An alternative way \cite{bar15:174107}
to perform TT+AS is to sample geometries before, and to minimize the
error between the approximation and the true values by coupling the
interpolation code and the quantum chemistry software. Avila and Carrington
\cite{avi15:044106} constructed the PES in the SOP form based on a
Smolyak interpolation technique using polynomial-like or spectral basis
functions and one-dimensional Lagrange-type functions. Habershon and
co-workers \cite{ric22:209} proposed an algorithm for tensor decomposition
of existing KMR prediction for on-the-fly MCTDH. Furthermore, like the SOP/CP-FBR
methods \cite{pan20:234110,nad23:114109} various approaches were proposed
combining tensorized Chebyshev interpolation with a function approximation
procedure. For instance, Dolgov and 
co-workers \cite{dol21:A2190} combined tensorized Chebyshev interpolation
with a Tucker decomposition of low multilinear rank to approximate a
three-dimensional function defined on a tensor-product domain via
function evaluations. Moreover, the functional tensor-train (FTT)
\cite{gor19:59} and its extended version (EFTT) \cite{str24:54}
formats were proposed to compress and work with multidimensional
functions on tensor product domains combining tensorized Chebyshev
interpolation. Both FTT and EFTT are continuous extensions of the
TT decomposition using a TT ansatz by replacing the
three-dimensional TT cores with univariate matrix-valued functions.

We now examine the complexity and rank of CPD-GPR/SVR in comparison
with SOP-NN \cite{man06:194105,koc14:021101}. As given in Table
\ref{tab:methods-sop-cpd}, the ranks of SOP-NN and CPD-GPR are equal
to the number of neurons and number of training data, respectively,
both of which typically exceed the number of support vectors (rank of
CPD-SVR) \cite{son22:1983,mia24:2410.23529}. The KMR approach offers
distinct advantages through its Bayesian framework, enabling PES
construction with reduced training data requirements 
\cite{kam18:241702,son19:114302,son20:134309,son22:1983}.
Thus, the rank of CPD-GPR often remains comparable to that of SOP-NN,
while CPD-SVR typically demonstrates lower rank due to $n_{\mathrm{sv}}<n$.
Consequently, CPD-GPR and SOP-NN exhibit similar levels of complexity
and nonlinearity, with CPD-SVR \cite{son22:1983,mia24:2410.23529}
occupying an intermediate position. However, the comparison becomes
more nuanced when considering ML-SOP-NN versus CPD-GPR/SVR. Extending
this analysis to ML-CPD-GPR/SVR reveals a consistent pattern where SVR
maintains its advantage in generating more compact expansion forms
compared to both NN and GPR approaches. From the view of tensor
decomposition, as indicated in Table \ref{tab:tensor-decomp-pes}, these
methods exhibit progressively lower implementation complexity in the
sequence from HT and Tucker decompositions to CP decomposition and to
TT decomposition. This sequence is consistent with the aforementioned
scenario if one notes that SOP-NN and CPD-GPR/SVR are based on Tucker
decomposition and their multilayer counterparts adopt HT decomposition.
The challenge of large-rank essentially arises from incomplete expansion
of the PES and bears conceptual similarity to the electron-correlation
problem in electronic structure theory. This analogy suggests both the
necessarity and feasibility of further improvements in building decomposed
PES. This laboratory has been actively developing two distinct approaches
to address this issue. The first approach leverages the inherent advantage
of PES reconstruction in achieving reduced ranks through optimized
expansion expression \cite{sch20:024108,men21:2702}. This has led to
the development of hybrid approach that combine direct construction
and reconstruction ways to effectively mitigate large-rank challenge.
The second innovative approach incorporates unsupervised learning
techniques, say the chemistry-informed generative adversarial network
(CI-GAN) method \cite{mia24:532} proposed in this laboratory. This
method \cite{mia24:532} generates distributions of geometry and energy
by $n\lesssim10^2$ training data, thereby offering the potential to
produce very small rank of $\sim10^2$.

%------wave function---------------
\section{Wave Function and Propagation\label{sec:working}}

\begin{figure}[h!]
  \centering
   \includegraphics[width=18cm]{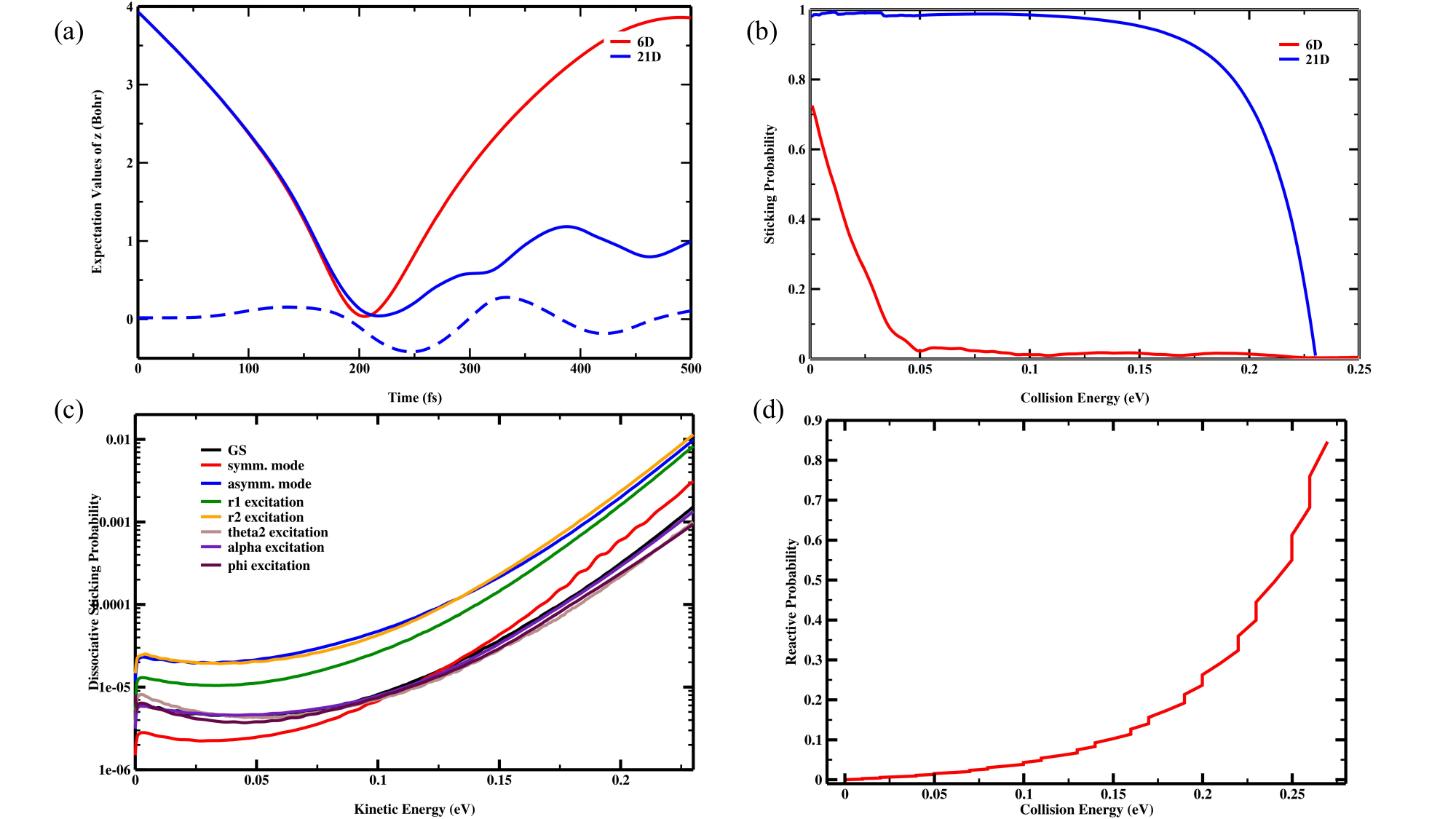}
    \caption{\figfoot}
     \label{fig:results-flux-exp}
      \end{figure}

\begin{figure}[h!]
  \centering
   \includegraphics[width=18cm]{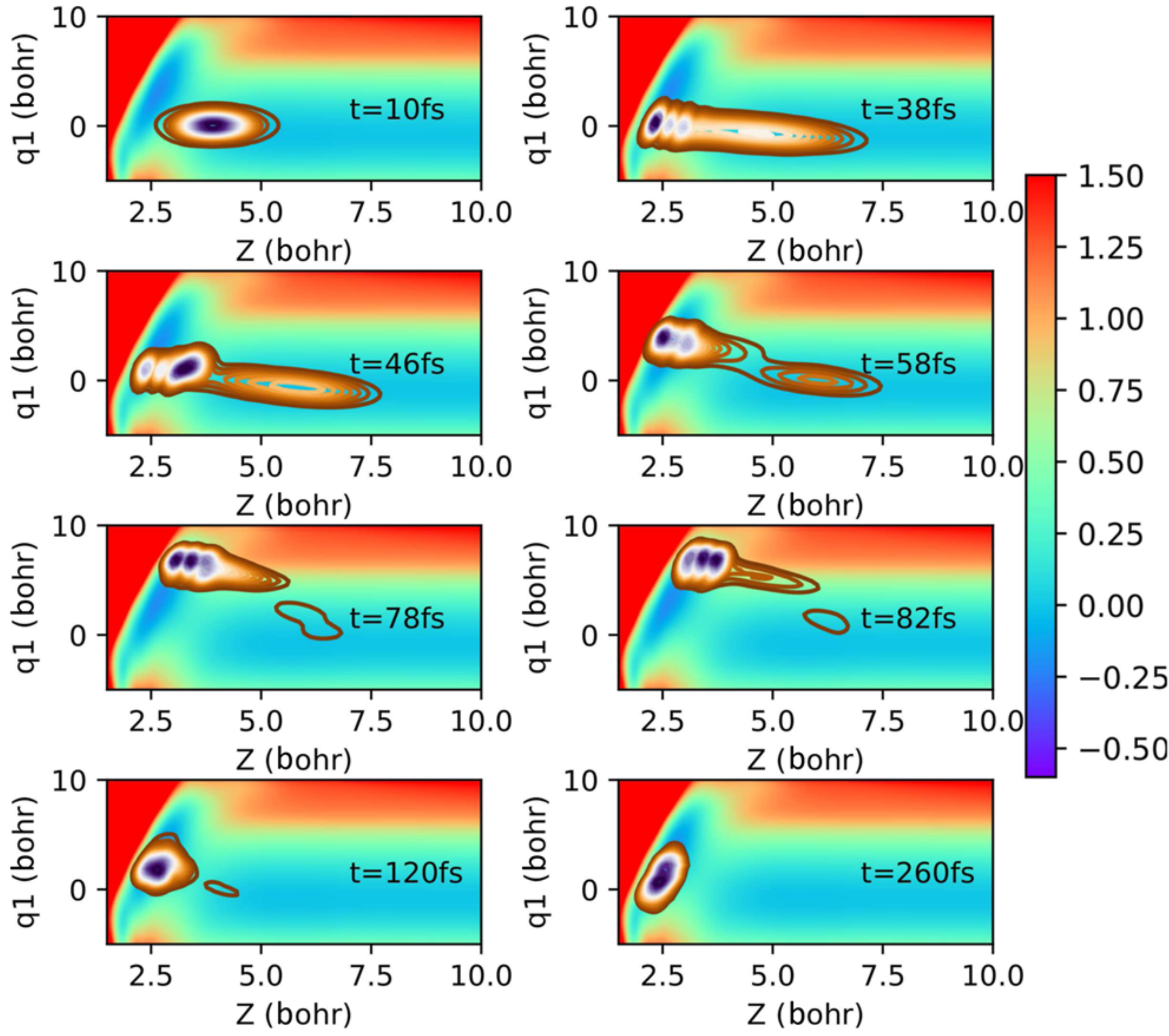}
    \caption{\figfoot}
     \label{fig:h-gra-dyna}
      \end{figure}

Now, let us consider the expression of nuclear wave function and
methods of wavepacket propagation. It has been described elsewhere
\cite{mey90:73,man92:3199,mey93:141,mey98:3011,bec00:1,mey03:251,vie03:3434,wu03:14,wan03:1289,wu04:2227,man08:164116,wan09:024114,ven11:044135,wod12:124119,mey12:351,men12:134302,men13:014313,men14:124309,wan15:7951,man17:253001,men17:184305,kur18:194114,son22:6047,dor24:8767,lar24:e2306881}
that the nuclear wave function can be expressed in a summation of
products of single-particle functions (SPFs) as shown in Equation
\eqref{eq:expansion-sop-form-01}. Repeatly expanding the SPFs, the wave
function is given in the multilayer expansion form. We refer the reader
to the review papers by Meyer and co-workers \cite{bec00:1,mey12:351},
Wang \cite{wan15:7951}, Manthe \cite{man17:253001}, and Larsson
\cite{lar24:e2306881} for details of this idea together with its
historical evolution. Specifically, the SPFs in the $(l-1)$-th layer
are expanded by products of another series of SPFs in the deeper $l$-th
layer 
\cite{wan03:1289,man08:164116,ven11:044135,mey12:351,wan15:7951,man17:253001,lar24:e2306881},
\begin{equation}
\varphi_m^{(\mathfrak{z}-1;\kappa_{l-1})}\Big(q^{(\mathfrak{z}-1;\kappa_{l-1})},t\Big)=
\sum_{j_1}^{n_{1}}\cdots\sum_{j_{p_{\kappa_l}}}^{n_{\kappa_l}}
A_{m;j_1,\cdots,j_{p_{\kappa_l}}}^{(\mathfrak{z})}(t)
\prod_{\kappa_l=1}^{p_{\kappa_l}}
\varphi_{j_{\kappa_l}}^{(\mathfrak{z},\kappa_l)}
\Big(q^{(\mathfrak{z};\kappa_l)}\Big)
=\sum_JA_{m;J}^{(\mathfrak{z})}\cdot\Phi_J^{(\mathfrak{z})},
\label{eq:mctdh-wavefunction-001}
\end{equation}
where $\mathfrak{z}=\{l;\kappa_1,\cdots,\kappa_{l-1}\}$ and
$\mathfrak{z}-1=\{l-1;\kappa_1,\cdots,\kappa_{l-2}\}$, while
$q^{(\mathfrak{z}-1;\kappa_{l-1})}=\{q^{(\mathfrak{z};1_{\kappa_l})},
\cdots,q^{(\mathfrak{z};p_{\kappa_l})}\}$ is a mode. The SPFs of the
last layer are finally expanded by a given set of primitive TI basis
functions. It is convenient to introduce a diagrammatic notation,
called ML-tree structure (see Figure \ref{fig:wf-ml-mctdh} for the
present three examples) to represent these objects
\cite{wan03:1289,man08:164116,ven11:044135}. By number of expansional
layers, one can classify propagation methods as collected in Table
\ref{tab:compare-nuclear-elec}. It is worth noting that wavepacket
propagation methods can be classified and named not only by their
expansion form but also by the time integrators
\cite{wan03:1289,man08:164116,ven11:044135,lub04:355,lub15:917,klo17:174107,bon18:252,lin21:174108,lin21:174109,wan18:044119}.
Next, by substituting the wave function and Hamiltonian operator in
the time-dependent variational principle ({\it i.e.}, Dirac-Frenkel
variational principle), Meyer, Manthe, Wang, and their co-workers
\cite{mey90:73,bec00:1,wan03:1289,man08:164116,ven11:044135,men12:134302}
derived the working equations for propgating SPFs and expansion
coefficients. Due to the equivalence of all layers, the working
equations for each layer are formally identical 
\cite{wan03:1289,man08:164116,ven11:044135}. This feature is particularly
advantageous for high-dimensional systems. Thus, by a given time-integrator
\cite{wan03:1289,man08:164116,ven11:044135,lub04:355,lub15:917,klo17:174107,bon18:252,lin21:174108,lin21:174109,wan18:044119},
the multi-layer expansion consequently provides a general framework
of propagation, enabling the independent propagation of SPFs. Using
the ML-MCTDH method, the present benchmarks have been extensively
investigated at this laboratory
\cite{son24:597,mia24:532,son22:11128,men21:2702,son22:6047} and the
corresponding numerical results are illustrated in Figure \ref{fig:results-flux-exp}.
We refer the reader to the Supporting Information for numerical details
of these ML-MCTDH calculations. Recently, Gatti and co-workers
\cite{shi23:194102,shi25:1896} reported their 75D ML-MCTDH calculations
for a hydrogen atom scattering from graphene, as illustrated in Figure
\ref{fig:h-gra-dyna}. These numerical results well demonstrate power
and capability of the ML-MCTDH method in computing quantum dynamic
properties of a molecular reaction, such as spectrum and reaction
probability.

One of the authors \cite{men12:134302,men13:014313} suggested the
necessity of optimizing the ML-tree structure to ensure efficient
convergence during propagation. As rules of thumb
\cite{men12:134302,men13:014313,men14:124309,men17:184305,men21:2702,son22:6047,son24:597},
the following points should be noted when designing an initial ML-tree
according to the hierarchical coordinate frames. First, two or three
coordinates from a single BF frame in the deepest layer should be
combined into one mode. Additionally, distance and angular coordinates
should be separated into distinct modes. Second, if more than three
coordinates are defined in a BF frame, they should be further divided
into separate BF frames in deeper layers. Modes belonging to the same
fragment may be grouped into a single mode in the upper layer. Third,
the three Euler angles of a relative frame should form a single mode.
Distance coordinates defining the reaction coordinate should be
separated into individual modes, each containing only one coordinate.
This separation is computationally advantageous as it reduces the
number of grids required along the reaction coordinate to achieve
well-converged propagation. The initial ML-tree is iteratively optimized
\cite{men12:134302,men13:014313} through extensive test calculations
where one should note that the larger the mode populations, the larger
are the couplings among the modes in the lower layers. For the present
benchmarks, the optimized ML-tree structures shown in Figure
\ref{fig:wf-ml-mctdh} differ from the hierarchical coordinate frames
shown in Figure \ref{fig:ml-expansion-examples}. This discrepancy is
expected as the coordinate frames are artificially designed. However,
a comparison between Figure \ref{fig:wf-ml-mctdh} and Figure
\ref{fig:ml-expansion-examples} reveals that most of the separations
are preserved suggesting physics inghts behind the hierarchical
coordinate frame. It should be mentioned that the above discrepancy
of separating schemes between the wave function and coordinates frame
decomposition does not affect the ML-MCTDH results. The tree structures
indicate separating scheme of the modes in either designing the
coordinates frame or expanding the wave function. The resulting
equations of motion (EOMs) of individual DOFs are essentially
independent on separating scheme. Therefore, the tree-structure
difference between coordinates frame and wave fucntion is acceptable
in solving the working equations for propagation. Later, Vendrell and
co-workers \cite{men23:1144} proposed an approach to design optimal
ML-tree structure using multivariate statistics, or more specifically,
factor analysis and hierarchical clustering. By the TN language,
Hikihara and co-workers \cite{hik23:013031} proposed a tree algorithm
to automatically optimize the network structure by local reconnections
of isometries to suppress the bipartite entanglement entropy on their
legs. This algorithm can be implemented to optimize the tree-structures
in DMRG.

We now address the question of whether the multi-layer expansion form
is sufficiently flexible for propagation. To this end, we compare the expansion form
of the nuclear wave function with that of the electronic wave function
\cite{sza96:book,wou14:272}, as summarized in Table \ref{tab:compare-nuclear-elec}.
We refer the reader to the Supporting Information for such comparison
details. Of course, we must mention that the motion of electrons and
nuclei during reactions is fundamentally different making electronic-structure
theory and wavepacket propagation methods physically distinct. In
general, the HF, full-CI, and MCSCF wave functions are expaned in
zero-, one-, and two-layer forms, respectively, while the
quantum chemistry TI-DMRG state is expanded in the multilayer form. We
compare the TI-DMRG or called QC-DMRG \cite{wou14:272,sza15:1342} with
ML-MCTDH using various criteria, as given in table \ref{tab:dmrg-alg-ml-mctdh}.
Moreover, numerical comparisons \cite{mai21:174106,lar22:11119,dor24:8767}
of ML-MCTDH with TD-DMRG have been reported by representing the state
and Hamiltonian in terms of MPSs and MPOs, respectively. Obviously,
these above approaches for either wavepacket propagation or electronic
structure employ the wave function in the CI type. In addition, let us
turn to the choice of CC-type nuclear wave function. Loosely, defining
reference state $\Phi_0$ of an $N$-electron system, the CI and CC wave
functions are given by,
\begin{equation}
\Psi_{\mathrm{CI}}=\left(1+\sum_{p=1}^N\hat{c}_p\right)\Phi_0,\quad
\Psi_{\mathrm{CC}}=\exp\big(\hat{t}\big)\Phi_0=
\exp\left(\sum_{p=1}^N\hat{t}_p\right)\Phi_0,
\label{eq:cc-ci-wave-func-00}
\end{equation}
where $\hat{c}_p$ and $\hat{t}_p$ introduce the $p$-excitation term,
while $N$ is the expansional order. Equation \eqref{eq:cc-ci-wave-func-00}
clearly demonstrates the approximate equivalence \cite{bar07:291}
between $\Psi_{\mathrm{CI}}$ and $\Psi_{\mathrm{CC}}$ for finite $N$.
The CC-type electronic wave function at $N\to\infty$ and the exact
solution of the time-dependent SE have equivalence formulation if the
nuclear Hamiltonian is time-independent. In this context, the
full-CI-type wave function expressed in terms of multilayer
expansion is sufficient for propagation. Despite this,
time-dependent coupled-cluster (TDCC) theory has been developed for
simulating laser-driven electronic dynamics in atoms and molecules and
for simulating molecular vibrational dynamics \cite{sve23:e1666}.
Although TDCC was developed for TD nucleon dynamics \cite{hoo78:2380,hoo79:1971}
and TD electronic-structure theory \cite{goi18:e1341,li20:9951,sve23:e1666},
it shows applicability in the field of quantum molecular dynamics.
Until now, we discussed nothing about perturbation theory, especially
Rayleigh-Schr{\"o}dinger perturbation theory (RSPT), where solutions
of variational methods serve as references for further perturbations.
When all configurations are considered in perturbation, RSPT evolves
into many-body perturbation theory (MBPT). The CC-type wave function
generally offers a very convenient resummation of MBPT diagrams offering
an infinite-order approximation in selected cluster operators \cite{bar07:291}.
Specially, if $\hat{t}_p$ is a connected cluster operator corresponding
to $p$-fold excitations, the CC and MBPT wave functions are approximately
equivalent. In this context, the TD perturbation theory (TDPT) is widely
employed to understand excitation in the interaction picture with TD
perturbation. However, for propagation with a TI Hamiltonian, separating
the Hamiltonian into a core term and a perturbation is often impossible,
making few, if any, perturbation theory applicable.

Nevertheless, CI-type, CC-type, and perturbative wave functions
necessarily emerge when solving the molecular rovibrational eigenstate
problems by the vibrational self-consistent field (VSCF) method and
its enhanced variants, as given in Table \ref{tab:comp-vibrational-modes-vscf}.
These methods variationally minimise a trial nuclear wave function
composed of a single Hartree product of one-dimensional vibrational
wave functions. Therefore, these vibrational-structure methods
align more closely with the paradigm of electronic-structure theory than
with that of dynamical methods which propagate the wave function in
imaginary time to reach the rovibrational eigenstates. Furthermore,
due to the wave function in the SOP form, these methods based on VSCF
also require the Hamiltonian operator in the similar form, like the
MCTDH and ML-MCTDH methods. We refer the reader to References
\cite{bow03:533,chr04:2140,pel10:20603,sei11:054119,fau15:134105,fau18:054104}
for VSCF and its variants.
Like the relation between MCTDH and TI-DMRG, the vibrational-structure
methods also have TI-DMRG counterparts \cite{gla23:9329}.
In addition to TI vibrational-structure problems, these methods can
be further extended to the TD propagations. For instance, in recent
decades, Christiansen and co-workers \cite{hoj24:024105,jen25:084112,hoj25:e70001}
developed coupled cluster theory for time-dependent wave functions for
the efficient computation of the quantum dynamics associated with the
motion of nuclei, including time-dependent vibrational coupled cluster
(TDVCC) and time-dependent modal vibrational coupled cluster (TDMVCC)
\cite{hoj24:024105,jen25:084112,hoj25:e70001}, which employ static and
adaptive basis sets, respectively.
At the present, developments of TDVCC and TDMVCC are still in their infancy.
The fast configuration-space convergence, flexible choice of basis type
and coordinate system, and polynomial-scaling computational cost make
TDVCC and TDMVCC promising \cite{hoj24:024105,jen25:084112,hoj25:e70001}.

%%%%%%%%%DISCUSS%%%%%%%%%%%%%%%%%%%%%%%%%%%
\section{Perspectives on Quantum Molecular Dynamics\label{sec:disscuss}}

\begin{figure}[h!]
  \centering
   \includegraphics[width=18cm]{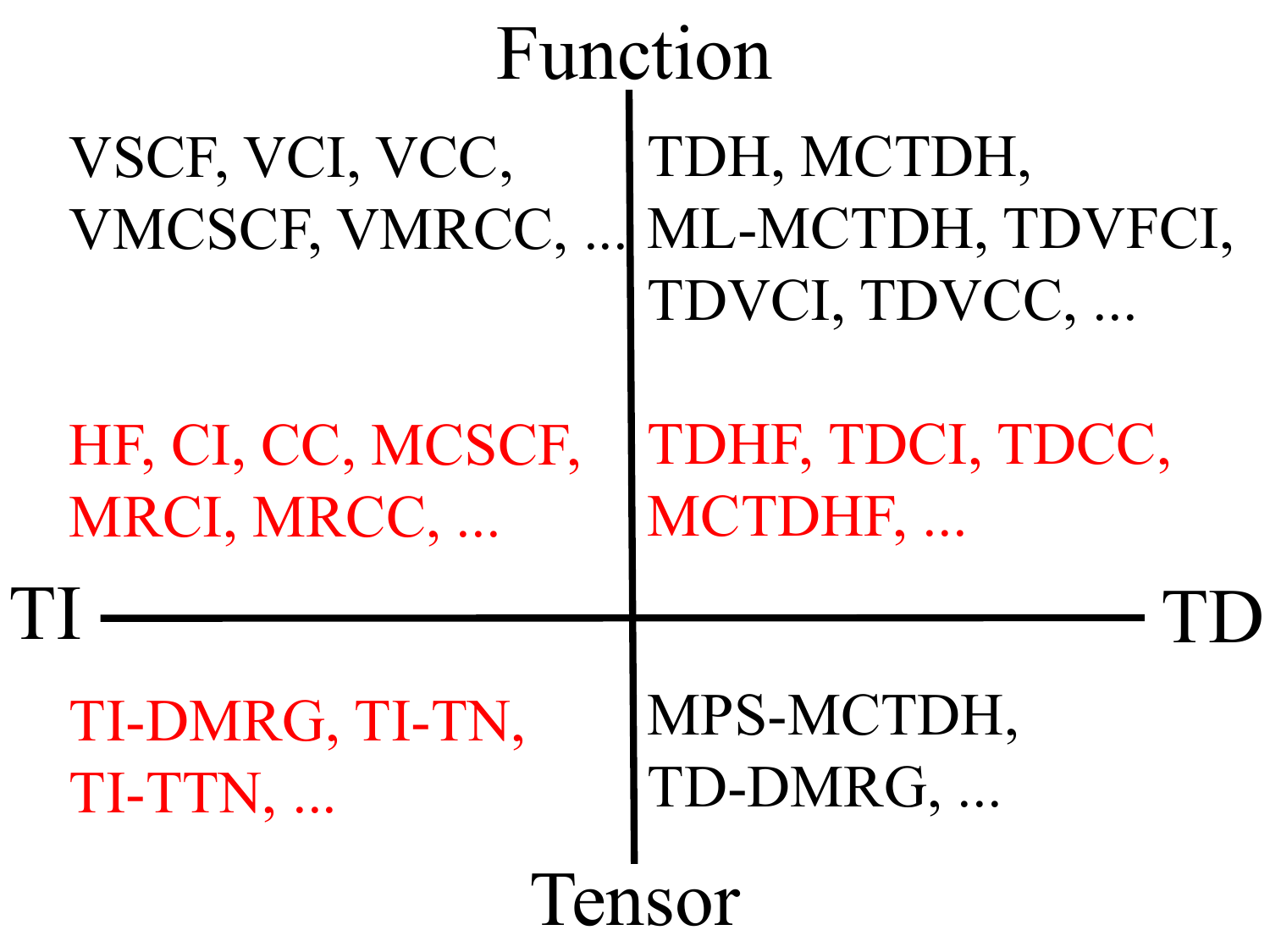}
    \caption{\figfoot}
     \label{fig:classif}
      \end{figure}

In the present section, let us discuss the classifications of the methods
for propagating wave function (as illustrated in Figure \ref{fig:classif})
and then give perspectives on quantum molecular dynamics. In Table
\ref{tab:comp-prop-methods}, we compare the propagation methods for
molecular reaction dynamcis with those for electronic and nucleon
dynamics. All of these methods were designed to propagate wave function
of either distinguishable particles (say atoms) or indistinguishable
particle (say electrons and nucleons). As indicated in Section
\ref{sec:working}, substituting a solution ansatz into the time-dependent
variational principle, one can derive the working equations. By solving
these working equations one can obtain resulting dynamics properties,
such as reaction probability, scattering cross section. In these methods,
the solution ansatz is expanded by products of one-dimensional (or
lower-dimensional) functions, called SPFs or MOs
in different scenarios. Correspondingly, the Hamiltonian operator is
expanded into a similar summation of products of one-dimensional (or
lower-dimensional) operators. This is helpful to reduce dimensionality
of the working equations because multi-dimensional integrators are
reduced to lower-dimensional ones. Repeatly expanding, we can finally
obtain a hierarchical framework for various dynamics scenarios. In
Table \ref{tab:comp-prop-methods}, we compare the hierarchical frameworks
for these scenarios. For instance, the time-dependent Hartree-Fock
(TDHF) wave function for either electron or nucleon has zero-layer
expansion form, similar to the standard propagation for chemical dynamics.
Similar to the HF eigensolver, TDHF is an impractical tool due to its
computational expense and lack of correlation effects making more
layers necessary. Consequently, the time-dependent version of CI, CC,
and density functional theory (DFT) were subsequently developed to
account for correlation effects leading to the TDCI, TDCC, and TDDFT
methods for electron or nucleon. Turning to the two-layer cases, the
MCTDH-type expansion with permutation asymmetry can be used to propagate
wave function of fermion leading to MCTDHF. In addition, there also
exists the MCTDHB method for boson. Therefore, if the hierarchical
expansion preserves the permutation symmetry or antisymmetry, it can be used as a
solution ansatz to derive a definitive propagation approach. In this
context, due to the symmetry, field-theory techniques can be used to
representation the wavepacket propagation making the second-quantization
representation play an important role in dynamics calculations.

Following Section \ref{sec:coord-keo} and the above discussions, we now
turn to the second-quantization representation (SQR) of quantum molecular
dynamics. This is a way to apply quantum field theory to molecular
dynamics, where the chemical dynamics is transferred to dynamics of
quantum field. The occupation number vector (ONV) specifies which and
how many SPFs are occupied by defining a vacuum state $\vert0\rangle$
alongside creation operator $a^{(\kappa)}_{i_{\kappa}}{}^{\dagger}$
and annihilation operator $a_{i_\kappa}^{(\kappa)}$ for the $\kappa$-th
coordinate together with commutators
\begin{equation}
\left[a_{i_\kappa}^{(\kappa)},a_{j_\eta}^{(\eta)}{}^{\dagger}\right]_{\pm}
=\delta_{\kappa\eta}\delta_{i_\kappa j_\eta},\quad
\left[a_{i_\kappa}^{(\kappa)},a_{j_\eta}^{(\eta)}\right]_{\pm}=
\left[a_{i_\kappa}^{(\kappa)}{}^{\dagger},a_{j_\eta}^{(\eta)}{}^{\dagger}\right]_{\pm}
=0,\quad\kappa,\eta=1,2,\cdots,d,\quad
i_\kappa,j_\eta=1,2,\cdots,
\label{eq:commutators-0000}
\end{equation}
where $+$ and $-$ mean anticommutators for fermions and commutators for
bosons, respectively.
If $\{a_{i_\kappa}^{(\kappa)},a_{i_\kappa}^{(\kappa)}{}^{\dagger}\}$
are defined for the $\kappa$-th mode, the SPFs in the same layer can
be expanded by products of states represented by acting a creation
operator on $\vert0\rangle$. Repeating this process, the quantum state is
finally represented in the multi-layer formalism. Since the
occupation and coordinate representations are equivalence, the SQ
Hamiltonian $\hat{\mathfrak{H}}$ has the same decomposed structure in
the matrix elements $\langle\Phi_I\vert\hat{H}\vert\Phi_J\rangle_{\mathrm{FQ}}
=\langle\Phi_I\vert\hat{\mathfrak{H}}\vert\Phi_J\rangle_{\mathrm{SQ}}$.
Such hierarchical SQR might be helpful in simulating many-body open
quantum systems \cite{ye25:120901} which have strong quantum correlations
in both space and time.

Over the past years, Wang and Thoss \cite{wan18:13}, Christiansen and co-workers \cite{mad18:134110},
Manthe and Weike \cite{man17:064117,wei20:034101}, and Sasmal and
Vendrell \cite{sas20:154110} reported their SQR for MCTDH/ML-MCTDH
(called MCTDH-SQR/ML-MCTDH-SQR) or other methods for computing
(i) trongly correlated electronic model \cite{wan18:13},
(ii) vibrational eigen-energy \cite{mad18:134110},
(iii) imaginary time propagation \cite{wei20:034101}, and
(iv) non-adiabatic spin dynamics \cite{sas20:154110}. Furthermore,
many developments on MCTDH-SQR/ML-MCTDH-SQR were reported and mainly
concerned on dynamics and properties of the condensed matter rather
than quantum molecular dynamics. This is because the occupation number
representation in second quantization naturally describes collective
behaviors of identical particle (see also Equation \eqref{eq:commutators-0000}). 
Employing the SQR, for instance, Christiansen and co-workers \cite{hoj25:e70001}
developed the time-dependent full vibrational configuration interaction
(TDFVCI), TDVCI[$n$], and TDVCC[$n$] methods where $n$ means number of
the Hartree products in the solution ansatz. Usually, convergence of
TDVCI[$n$] is slower compared to TDVCC[$n$], highlighting the advantage
of the non-linear CC parameterization. We refer the reader to Reference
\cite{hoj25:e70001} for details of TDVCI and TDVCC under the SQR and to
Figure \ref{fig:classif} for comparison of TDVCI and TDVCC with other
metbhods. In addition to the above field-theory techniques for paticles,
as given in Sections \ref{sec:pes-cpd} and \ref{sec:working}, another
way to implement quantum molecular dynamics for large system
\cite{mai21:174106,lar22:11119,dor24:8767}, such as material simulations,
focuses on the methods with MPS and MPO, which is a fundamental form of
TN or TTN and the most widely used representation for quantum many-body
theory of condensed matter systems. In general, employing the grid-based
representation the wave function and Hamiltonian operator in the decomposition
form are represented by tensor decomposition. In this context, the methods
with MPS and MPO should be equivalent to those with functional forms,
as derived by Larsson \cite{lar24:e2306881}. A popular method with MPS
and MPO is the TD-DMRG method for propagating nuclear wavepacket, while
another approach is MPS-MCTDH \cite{kur18:194114,hin24:3839}. Over the
past decade, various comparisons \cite{mai21:174106,lar22:11119,dor24:8767}
of such methods with ML-MCTDH were reported by modeling vibrational or
photoelectron spectrum because the DMRG method was originally developed
to address challenges in simulating strongly correlated many-body systems.
The TD-DMRG method was found to be highly accurate for dynamics of
one-dimensional chain systems \cite{mai21:174106} and has been extended
to higher-dimensional models \cite{lar22:11119,dor24:8767}.

\begin{figure}[h!]
  \centering
   \includegraphics[width=18cm]{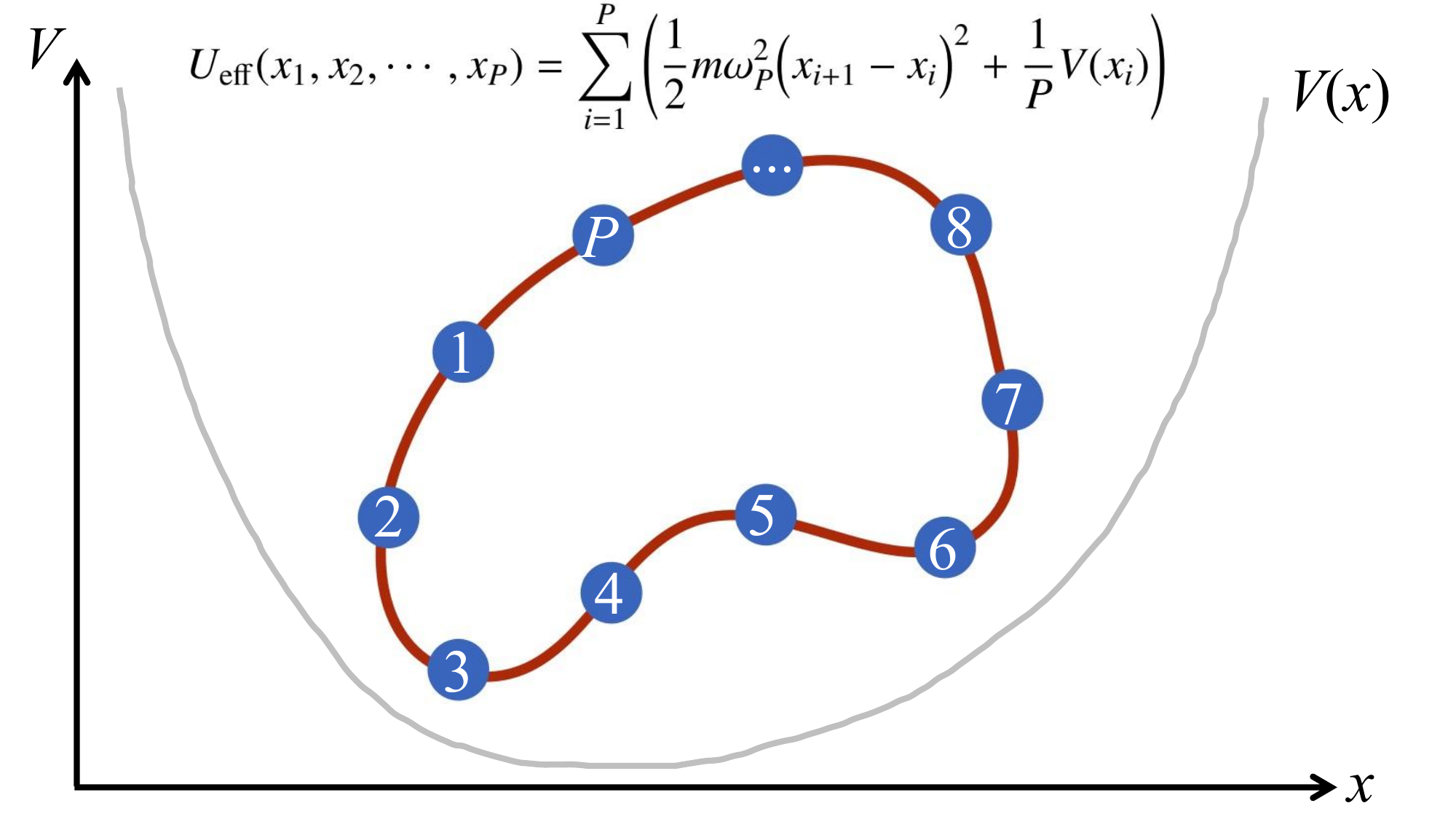}
    \caption{\figfoot}
     \label{fig:ring-polymer-ueff}
      \end{figure}

The concepts of ONV, creation, and annihilation operators introduce
virtual particle as spacial points along paths in a path-integral
framework. This demonstrates an equivalence between quantum dynamics
and path-integral molecular dynamics (PIMD)
\cite{sul11:044131,all13:221103,men15:101102,men16:154312,men16:29286,li23:5087}.
It is a classical dynamics formulation in imaginary-time mathematically
equivalent to a Wick rotation of path-integral quantum dynamics. The
PIMD method enables the simulation of temperature-dependent chemical
kinetics, extending beyond classical molecular dynamics to capture
quantum statistical effects. To understand this point, for a one-dimension
system with action $S=m\dot{x}^2/2-V(x)$ and Hamiltonian
$\hat{H}_{\mathrm{ori}}=-\partial^2_x/(2m)+V(x)$, the propagator is given by
\begin{equation}
K(x,x',t)=\Big\langle x\Big\vert\exp\Big(-i\hat{H}_{\mathrm{ori}}t\Big)\Big\vert
x'\Big\rangle
=\lim_{P\to\infty,\epsilon\to0}\left(\frac{m}{2i\pi\epsilon}
\right)^{\frac{P}{2}}
\int_x^{x'}\exp\Big(iS\Big)\mathrm{d}x_1\cdots\mathrm{d}x_P,
\label{eq:path-integral-for-000-pimd}
\end{equation}
where the path from $x$ to $x'$ is divided into $P+1$ parts by points
$\{x_i\}_{i=1}^P$, and the time from $0$ to $t$ is divided into
$\epsilon+1$ parts such that $P\epsilon=t$. By introducing imaginary time
$\beta=it$, interpreted as inverse temperature, the partition function becomes
\begin{align}
Z(\beta)={}&\mathrm{Tr}\Big\langle x\Big\vert\exp(-\beta\hat{H}_{\mathrm{ori}})
\Big\vert x\Big\rangle
=\lim_{P\to\infty}\left(\frac{mP}{2\pi\beta}\right)^{\frac{P}{2}}
\int\exp\left[-\sum_{i=1}^P\left(\frac{mP}{2\beta}
\Big(x_{i+1}-x_i\Big)^2+\frac{\beta}{P}V(x_i)\right)\right]
\mathrm{d}x_1\cdots\mathrm{d}x_P
\allowdisplaybreaks[4] \nonumber \\
\propto{}&\lim_{P\to\infty}\int\exp\Big(-\beta\hat{H}_{\mathrm{eff}}^{(P)}\Big)
\mathrm{d}x_1\cdots\mathrm{d}x_P
\mathrm{d}p_1\cdots\mathrm{d}p_P,
\label{eq:path-integral-for-003-pimd}
\end{align}
yielding the effective Hamiltonian, known as the ring-polymer Hamiltonian,
\begin{equation}
\hat{H}_{\mathrm{eff}}^{(P)}=\sum_{i=1}^P\left[\frac{\hat{p}_i^2}{2m}
+\frac{1}{P}V(x_i)+\frac{1}{2}m\omega_P^2\Big(x_{i+1}-x_i\Big)^2\right],
\quad2\omega_P^2=mP/\beta^2.
\label{eq:path-integral-for-004-pimd}
\end{equation}
The ring-polymer Hamiltonian $\hat{H}_{\mathrm{eff}}^{(P)}$ describes
an ensemble of $P$ replicas of original system where two adjacent replicas
interact through a harmonic-oscillator potential with frequency $\omega_P$,
as shown by Figure \ref{fig:ring-polymer-ueff}.
Over the past decades, power of ring-polymer molecular dynamics (RPMD)
in simulating kinetics has been demonstrated by this laboratory and other groups
\cite{sul11:044131,all13:221103,men15:101102,men16:154312,men16:29286,li23:5087}.
When $P\to\infty$, Equations \eqref{eq:path-integral-for-000-pimd}
and \eqref{eq:path-integral-for-003-pimd} indicate equivalence
between $d$-dimensional quantum system and $(d+1)$-dimensional
classical system, referred to as an equivalent classical system.
However, we must mention that, this terminology is appropriate only
in cases where all the path weights are positive. It will be true for
systems which quantum Monte Carlo (QMC) calculations can be performed
(see also Equation \eqref{eq:path-integral-for-003-pimd}) in practical
implementation. In some cases, the $(d+1)$-dimension effective model
is an anisotropic analog of the classical model in $d+1$ dimensions.
In most cases, the equivalent classical system does not simply
correspond to the original model in one higher dimension (comparing
$\hat{H}_{\mathrm{ori}}$ with $\hat{H}_{\mathrm{eff}}^{(P)}$). Rather,
the path integral typically maps to a fundamentally different statistical
mechanics problem in $d+1$ dimensions, bearing no obvious resemblance
to the original $d$-dimensional quantum system. In the last case, the
$d$-dimensional quantum system is equivalent, on large length and time
scales, to the same classical system in $d+1$ dimensions.

Finally, we would like mention the new development on application of
generative artificial intelligence (AI) models to dynamics simulations.
A typical example is the CI-GAN method \cite{mia24:532}. In Section
\ref{sec:pes-cpd} we mentioned that CI-GAN has capability to directly
build the PES in the CPD form with smaller rank. In this case, the
training database only contains geometries and associated energies
without any time-dependent parameters. However, if the training data
are computed in time-series order, the generated geometries are, in
principle, in the time-series order. In other words, if the training
data contain time or time-dependent quantity (say trajectory or wave
function) it is possible to generate its time for each generated quantity
(say geometry or wave function). With time and associated geometry one
can actually obtain a trajectory. We would like to emphasize that this
is not a simple and direct extension of the CI-GAN method \cite{mia24:532}
because the time index is certainly different from geometries and associated
energies. Setting the beginning time to be zero, $t_0=0$, based on geometry
$\mathbf{X}(t_0)$ as well as its small neighbourhood, one can generate
geometries $\mathbf{X}(t)$ by CI-GAN. As shown above, these time-dependent
geometries are expected to replace a classical trajectory. But, the
predicted distributions of both time and geometry still require further
developments and considerations. Very recently, we derived a generator
for generation of the time index. However, if we turn to the time-dependent
wave function, it is far more complex than the above discussions and the
CI-GAN method \cite{mia24:532}. One can expect that it is very difficult
in implementation for generating quantum dynamics.

%%%%%%%%%%%%%%%%%%%%%%%%%%%%%%%%%%%%%%%%%%%%%%%%%
\section{Conclusions\label{sec:conclud}}

In this work, approaches to propagate nuclear wave function are reviewed
though a hierarchical framework under the coordinate representation.
Its core concept is how to combine several coupled coordinates into a
single mode making the coordinates defined in hierarchical frame. By
the hierarchical coordinate frame, the Hamiltonian operator and wave
function are expanded by products of single-particle operators and
functions, respectively. By substituting the multilayer solution ansatz
to the Dirac-Frenkel variational principle, a set of coupled working
equations for the expansion coefficients and the single-particle functions
can be derived. First, the coordinate frame is hierarchically designed
enabling the KEO to be expressed in a SOP form. Defining the coordinate
frame and interested reaction coordinate, one can sample geometries
based on which electronic-structure energy calculations are performed
to compute training dataset. Second, the PES is decomposed by two ways,
direct construction by training dataset and reconstrcution through
existing PES. Thus, the Hamiltonian operator has been given in summation
of products of single-particle operator. Third, the wave function is
expanded repeatedly by a series of configurations which are products
of SPFs. Since the Hamiltonian operator and wave function are both
expressed in multilayer summation form, the working equations can be
proved to be identical for all layers. This implies that the present
hierarchical framework is useful for propagating high-dimension wave
function. To show this point, the present benchmarks extensively
investigated at this laboratory \cite{son24:597,mia24:532,son22:11128,men21:2702,son22:6047}
by the ML-MCTDH method. The numerical results of CO/Cu(100), H$_2$O/Cu(111),
and HO$_x$ are illustrated in Figure \ref{fig:results-flux-exp}.
Moreover, 75D ML-MCTDH calculations for a hydrogen atom scattering from
graphene \cite{shi23:194102,shi25:1896} are also illustrated in Figure
\ref{fig:h-gra-dyna} well demonstrating power and capability of the
ML-MCTDH method. In addition, advantages and disadvantages of the present
hierarchical framework are discussed to give perspectives on molecular
reaction dynamics.

%-------------------------------
\section*{Acknowledgements}

The authors gratefully acknowledge financial support by National
Natural Science Foundation of China (Grant No. 22273074)
the {\it Centre national de la recherche scientifique} (CNRS)
International Research Network (IRN) ``MCTDH''. The authors
are grateful to Prof. Dr. H.-D. Meyer and Dr. M. Schr{\"o}der
(Universit{\"a}t Heidelberg), Prof. Dr. F. Gatti and Prof. Dr. D.
Pel{\'a}ez (Universit{\'e} Paris-Saclay) for helpful discussions and
also grateful to anonymous reviewers for critical discussions.

%--------------------------------
\section*{Supporting Information}

The Supporting Information is available free of charge at http://dx.doi.org/XXX.
This file contains (1) implementation details of the present multi-layer unified theory
on the benchmarks of CO/Cu(100), H$_2$O/Cu(111), and HO$_x$; (2) details
of the approaches to build the PES in general form, (3) details of the
SOP-NN, ML-SOP-NN, CPD-GPR/SVR and ML-CPD-GPR/SVR methods, and (4) detail
perspectives on (i) comparisons to electronic-structure approaches, (ii)
Fourier transformation, (iii) multi-layer SQR and PIMD, and (iv) the RPMD
method.

%%%%%%%%%%%%%%%%%%%%%%%%%%%%%%%%%%%%%%%%
% TABLES
%%%%%%%%%%%%%%%%%%%%%%%%%%%%%%%%%%%%%%%%
%%%%tab.1%%%%%%%%%%%%%%%%%%%%%%%
\clearpage
\begin{table}
\caption{%
Comparison of various methods to build decomposed PES with the MCTDH
language, including direct construction (upper panel) and reconstruction
(lower panel) methods. Abbreviations of these methods are explained in
the main text. The third column gives number of expansional layers of
potential function. The fourth column gives techniques to derive these
approaches. The fifth and sixth columns give rank of the resulting PES
and remarks on these methods, respectively. The rightmost column gives
references that reported these methods.
}%
\begin{tabular}{llllllllllllr}
\hline
No. & Method &~~& Layer &~~& Technique &~~& Rank &~~& Remark &~~& Reference \\ 
\hline
\multicolumn{12}{l}{{\it Direct Construction}} \\
1 & SOP-NN \footnote{The SOP-NN function of Equation \eqref{eq:nn-sop-form-000}
is actually in the CPD form.}
    && one   && GLR/NN && $F^{(1)}$  &&
rank is number of neurons
%Equation \eqref{eq:nn-sop-form-000} with one-hidden-layer 
&& References \cite{man06:194105,koc14:021101}   \\
2 & ML-SOP-NN && multi && GLR/NN && $\sum_{l}F^{(l)}$ &&
rank is number of neurons
%Equation \eqref{eq:km-sop-prediction-88} with multi-hidden-layer  
&& this work 
\footnote{We refer the reader to the Supporting Information
for details.\label{foot:multi-layer-pes-cpd}} \\
3 & CPD-GPR  && one && KMR/GPR && $n$
\footnote{Values of $n$ and $n_{\mathrm{sv}}$ are numbers of training data and
support vectors satifying $n_{\mathrm{sv}}<n$.\label{foot:rank-cpd-gpr}}
&& $n$ cannot be optimized && Reference \cite{son22:11128} \\
4 & CPD-mc-GPR && one && KMR/GPR && $n$ \textsuperscript{\ref{foot:rank-cpd-gpr}} 
&& mode combination is adopted && Reference \cite{son24:597} \\
5 & CPD-SVR && one && KMR/SVR && $n_{\mathrm{sv}}$ \textsuperscript{\ref{foot:rank-cpd-gpr}}
&& $n_{\mathrm{sv}}$ can be optimized && Reference \cite{mia24:2410.23529} \\
6 & CPD-ws-SVR \footnote{It is also possible to extend CPD-ws-SVR to
its mode-combination version.}
&& one && KMR/SVR && $n_{\mathrm{sv}}$ \textsuperscript{\ref{foot:rank-cpd-gpr}}
&& warm-started SVR is used && Reference \cite{mia24:2410.23529} \\
7 & ML-CPD-GPR
\footnote{It can be extended to the mode-combination or warm-started version.
\label{foot:ml-cpd-gpr-svr-00}}
&& multi && KMR && $\sim l\cdot n$ \textsuperscript{\ref{foot:rank-cpd-gpr}}
&& multi-layer CPD-GPR && this work \textsuperscript{\ref{foot:multi-layer-pes-cpd}} \\
8 & ML-CPD-SVR \textsuperscript{\ref{foot:ml-cpd-gpr-svr-00}}
&& multi && KMR && $\sim l\cdot n_{\mathrm{sv}}$ \textsuperscript{\ref{foot:rank-cpd-gpr}}
&& multi-layer CPD-SVR && this work \textsuperscript{\ref{foot:multi-layer-pes-cpd}} \\
9 & RKHSI && one && interpolation && $n$ \textsuperscript{\ref{foot:rank-cpd-gpr}}
&& interpolation with kernel function && Reference \cite{hol97:7223}  \\
10 & SOP-FBR \footnote{Using FBR, the wave function is represented by a
finite set of time-independent functions.\label{foot:cp-sop-fbr}}
&& one && interpolation && $\sum_{\kappa=1}^dm_{\kappa}$
\footnote{Parameters $\{m_{\kappa}\}_{\kappa=1}^d$ are number of terms
in decomposed function.\label{foot:para-terms-expand}} && 
%SPPs are Chebyshev expanded
see $V^{(\mathrm{SOP-FBR})}$ in Equation \eqref{eq:sop-fbr-000}
&& Reference \cite{pan20:234110} \\
11 & CP-FBR \textsuperscript{\ref{foot:cp-sop-fbr}}
&& one && interpolation && $R$
\footnote{Number of expansion terms should be given before interpolation by
$V^{(\mathrm{CP-FBR})}$.\label{foot:rank-cpd}}  &&
%SPPs are Chebyshev expanded
see $V^{(\mathrm{CP-FBR})}$ in Equation \eqref{eq:sop-fbr-001}
&& Reference \cite{nad23:114109}  \\
\hline
\multicolumn{12}{l}{{\it Re-Fitting Methods}} \\
12 & POTFIT && one && regression &&
$\sum_{\kappa=1}^dm_{\kappa}$ \textsuperscript{\ref{foot:para-terms-expand}}
&& error as target for $d\leq7$ systems
&& References \cite{jae95:5605,jae96:7974,jae98:3772}  \\
13 & MGPF   && one && interpolation &&
$\sum_{\kappa=1}^dm_{\kappa}$ \textsuperscript{\ref{foot:para-terms-expand}}
&& interpolation from coarse to fine grids
&& Reference \cite{pel13:014108}  \\
14 & MLPF   && multi&& regression &&
$\sim l\cdot\sum_{\kappa=1}^dm_{\kappa}$ \textsuperscript{\ref{foot:para-terms-expand}}
&& suitable for $d>7$
&& Reference \cite{ott14:014106}  \\
15 & MCPF   && one && regression &&
$\sum_{\kappa=1}^dm_{\kappa}$ \textsuperscript{\ref{foot:para-terms-expand}}
&& Monte Carlo for $d>7$ SOP
&& Reference \cite{sch17:064105}  \\
16 & MCCPD  && one && regression && $R$
&& Monte Carlo for $d>7$ CPD
&& References \cite{sch20:024108,men21:2702}  \\   
\hline
\end{tabular}
\label{tab:methods-sop-cpd}
\end{table}

%tab.2%%%%%%%%%%%%%%%%%%%%%%%%%%
\clearpage
\begin{sidewaystable}
\caption{%
Comparison of various tensor-decomposition methods for approximating
function by tensor language, where an existing function is required.
For the PES construction, these methods belong to reconstruction scheme.
Abbreviations of these methods are explained in the main text. The third
column gives implementation complexity. The fourth column gives number of
expansional layers of potential function. The fifth column gives techniques
to derive these approaches. The sixth and seventh columns give rank of
the resulting PES and remarks on these methods, respectively. The rightmost
column gives references that reported these methods.
}%
\begin{tabular}{llllllllllllr}
\hline
No. & Method & Complexity & Layer &~~& Technique &~~& Rank &~~& Remark &~~& Reference \\ 
\hline
1 & CP     & medium & one && a sum of rank-1 tensors && low
\footnote{Rank of the CP decomposition is hardly determined.}
&& moderate dimensionality && References \cite{car70:283,kol09:455} \\
2 & Tucker & high & one && SVD to higher dimensions
\footnote{SVD means singular value decomposition.} 
&& --- \footnote{Rank of the Tucker decomposition is not clearly defined.\label{foot:tucker-rank}}
&& low dimensionality
\footnote{The Tucker decomposition requires that dimensionality is smaller than six.} 
&& Reference \cite{tuc66:279,kol09:455} \\
3 & FT   & high & one && combines with Chebyshev
&& --- \textsuperscript{\ref{foot:tucker-rank}} && three dimension && Reference \cite{dol21:A2190} \\ 
  &      &      &     && basis expansions && && &&  \\
4 & HT   & very high & multi && factorization using recursive  && --- && suitable for very 
&& References \cite{hac09:706,lub13:470} \\ 
  & && && subspace splitting && && high-dimensional function &&  \\ 
5 & TT    & low & one && a sequence of matrix products && finite && suitable for very
&& Reference \cite{ose10:70} \\
  & && && && && high-dimensional function &&  \\ 
6 & TT+AS \footnote{TT+AS is the first application of the TT cross
approximation (CA) procedure for the PES construction.}
& not high & one && an affine transformation of &&
finite \footnote{Rank should be determined carefully.\label{foot:rank-choose}} 
&& coordinates should be carefully && Reference \cite{bar15:174107}  \\
 &  &&  && Cartesian coordinates into &&  && determined to get a good && \\
 &  &&  && the active subspaces
\footnote{In the active subspaces (AS), the potential function has the most variability. The affine transformation is often used in the construction of the normal modes.}
&&  && low-rank tensor approximation && \\
 & &&  && && && of multidimensional function &&  \\
7 & FTT & not high & one && combines with basis expansions
&& finite \textsuperscript{\ref{foot:rank-choose}}
&& Chebyshev basis is a choice && Reference \cite{gor19:59} \\
8 & EFTT & not high & one && compressed FTT approximation
&& low && functional low-rank approximations && Reference \cite{str24:54} \\
9 & CA \footnote{CA means cross approximation, also known as skeleton decomposition.}
& medium & one && interpolation of selected tensor && finite &&
suitable for function with && Reference \cite{gor08:404} \\
  &  &&  &&  &&  &&  any dimensionality &&  \\  
\hline
\end{tabular}
\label{tab:tensor-decomp-pes}
\end{sidewaystable}

%%tab.3%%%%%%%%%%%%%%%%%%%%%%%%%%%%%%%%%%%%%%%%%%%%%%%%
\clearpage
\begin{table}
\caption{%
Comparison of the PES construction methods with the tensor decomposition
approaches for fitting the potential tensor in the sum of direct products
of one-dimensional tensors. The former is given in Table \ref{tab:methods-sop-cpd},
while the latter is given in Table \ref{tab:tensor-decomp-pes}. Since
we focus exclusively on the construction method irrespective of data
provenance, no fitting-type distinction of re-construction versus direct
construction is required here. Abbreviations of these methods are explained
in the main text. The first column gives methods of the tensor decomposition.
The second column gives correspondence methods for approximating the
decomposed PES. The third column gives expansional expression of
corresponding methods. The rightmost column gives remarks on these
methods, in particular rank of the expanded PES.
}%
\begin{tabular}{llllllr}
\hline
Tensor decomposition &~~& Function decomposition &~~& Expression &~~& Remark \\ 
\hline
Tucker && POTFIT && $V^{(\mathrm{SOP})}$ in Equation \eqref{eq:cpd-sop-form-4-pes}
&& POTFIT was designed for MCTDH  \\
      && MGPF &&  && MGPF is multi-grid variant of POTFIT
\footnote{MGPF and MCPF were designed for re-fitting
high-dimensional (larger than six) PESs to the SOP form.\label{foot:var-potfit}}   \\
      && MCPF &&  && MCPF is variant of POTFIT \textsuperscript{\ref{foot:var-potfit}} \\
HT     && MLPF && && MLPF is a multilayer variant of POTFIT 
\footnote{MLPF was designed for re-fitting
high-dimensional (larger than six) PESs to the multilayer SOP form.} \\
       && ML-SOP-NN      && && Rank of HT is not clearly defined \\
       && ML-CPD-GPR/SVR
\footnote{It means all variants of the CPD-GPR and CPD-SVR methods
given in Table \ref{tab:methods-sop-cpd}.\label{foot:cpd-gpr-svr-998}}
       && && \\
CP     && MCCPD     && $V^{(\mathrm{CPD})}$ in Equation \eqref{eq:cpd-sop-form-4-pes}
&& Rank should be carefully determined \\
&& SOP-NN  && Equation \eqref{eq:nn-sop-form-000}  &&  \\
&& CPD-GPR/SVR \textsuperscript{\ref{foot:cpd-gpr-svr-998}}
&& Equations \eqref{eq:cpd-gpr-4cpd} and \eqref{eq:cpd-gpr-4cpd-000}  && \\
FTT, EFTT && SOP-FBR && Equation \eqref{eq:sop-fbr-000}
&& Chebyshev basis is used to expand  \\
 && CP-FBR  && Equation \eqref{eq:sop-fbr-001}
&& the SPPs in $V^{(\mathrm{SOP})}$ or $V^{(\mathrm{CPD})}$ \\
\hline
\end{tabular}
\label{tab:comp-tensor-function-99}
\end{table}

%-tab.4-----------------propagation------------------
\clearpage
\begin{sidewaystable}
\caption{%
Comparison of wave functions between wavepacket propagation and
electronic structure theory. The left and right panels give properties
of the propagation and electronic-structure methods, respectively. The
leftmost column gives number of layers of expanded wave function. In
each panel, the first column gives symbols, while the other columns give
coefficient, configuration, and permutation symmetry requirement of the
entire wave function. The TI basis functions for propagation are also
indicated here, while those for electronic wave function is omitted.
}%
\begin{tabular}{lllllllllllllllll}
\hline
Layer \footnote{
Number of layer represents the number of expansions in building the
configuration.}
&~~&
\multicolumn{7}{c}{Wave function propagation}&~~&
\multicolumn{7}{c}{Electronic structure theory}  \\\cline{3-9}\cline{11-17}
&~~&Symbol \footnote{
The symbol in the form of A/B represents theoretical approach A and
the underingly basis set B.}
&~~&Coeff.&~~&Conf. \footnote{
We give the functions whose product consists of the configuration
which is product of a series of one-dimensional functions.\label{foot:propaga-c}}
&~~&Sym.&~~&Symbol&~~&Coeff.&~~&Conf. \textsuperscript{\ref{foot:propaga-c}}&~~&Sym. \\
\hline
zero \footnote{
The zero-layer expansion means the configuration itself is product
of the basis functions.}
          &~~& TDH/FBR      &~~& TD function      &~~& TI basis  &~~& no
          &~~& HF           &~~& need optimation  &~~& basis     &~~& antisymm  \\
one       &~~& standard/FBR &~~& TD function      &~~& TI SPF    &~~& no
          &~~& full-CI      &~~& need optimation  &~~& MO        &~~& antisymm  \\
%          &~~& TDCC        &~~& TD function      &~~& TI SPF    &~~& no
%          &~~& CC           &~~& need optimation  &~~& MO        &~~& asymm \\           
two       &~~& MCTDH/DVR
\footnote{The basis set of discrete variable representation (DVR)
%or correlation discrete variable representation (CDVR) 
is adopted.\label{foot:dvr-func}}
                            &~~& TD function      &~~& TD SPF    &~~& no
          &~~& MCSCF        &~~& need optimation  &~~& need optimation &~~& antisymm \\
%          &~~& %MCTDH/CDVR \textsuperscript{\ref{foot:dvr-func}}
%                            &~~& TD function      &~~& TD SPF and reference &~~& no
%          &~~& MCSCF-MBPT \footnote{Hybrid method with many-body perturbation
%theory (MBPT) is adopted.\label{foot:hybrid-mbpt}}   
%                            &~~& need optimation  &~~& need optimation/perturbation &~~& asymm \\
multi     &~~& ML-MCTDH/DVR \textsuperscript{\ref{foot:dvr-func}}
                            &~~& TD function      &~~& TD multi-layer SPF &~~& no
          &~~& TI-DMRG      &~~& ---              &~~& ---                &~~& antisymm
\footnote{The DMRG is typically performed in Fock space. Thus, its
antisymmetry is taken care of by the operators and not by the wavefunction.} \\ 
%          &~~& %ML-MCTDH/CDVR \textsuperscript{\ref{foot:dvr-func}}
%                            &~~& TD function      &~~& TD SPF and reference &~~& no
%          &~~& TI-DMRG-MBPT \textsuperscript{\ref{foot:hybrid-mbpt}}&~~& ---  &~~& ---         &~~& asymm \\
\hline
\end{tabular}
\label{tab:compare-nuclear-elec}
\end{sidewaystable} 

%-tab.5--------------------------------
\clearpage
\begin{sidewaystable}
\caption{%
Comparison and summary of TI-DMRG and ML-MCTDH. The second column
gives criteria to compare TI-DMRG with ML-MCTDH, including expansional
forms of Hamiltonian operator and wave function, optimization
approaches, time integrators, system features, and application
scenarios. Here, the first- and second-generation TI-DMRG methods
are given, denoted by o-DMRG and s-DMRG, respectively. Since the
DMRG theory was originally developed for field theory of many-body
problems (say condensed matter systems), o-DMRG employs the renormalized
or complementary Hamiltonian operators. In field theory, renormalized
operators focus on regularization and finiteness, while complementary
operators focus on symmetry and duality structures. In condensed matter
systems, renormalized operators and complementary operators are generally
distinct, but their mathematical expressions may overlap in high-symmetry
systems.
}%
\begin{tabular}{lllll}
\hline
No. & Criteria & \multicolumn{2}{c}{TI-DMRG} & ML-MCTDH  \\ \cline{3-4}
    &          & o-DMRG \footnote{Symbol ``o-DMRG'' means the original TI-DMRG theory, 
or frst-generation formulation.}
               & s-DMRG
\footnote{Symbol ``s-DMRG'' means the second-generation TI-DMRG theory defined by MPS and MPO.} & \\ 
\hline
1 & Hamiltonian & Renormalized or complementary & MPO & SOP or CPD 
\footnote{Only adopted by MCTDH/DVR. MCTDH/CDVR is not restricted by SOP or CPD form.} \\
2 & Wave function & MPS \footnote{Sometimes called it transformation matrices.}
& MPS \footnote{The MPSs are named as tensor train (TT)
factorization from the view point of numerical analysis.} & Multi-layer expansion  \\
3 & Optimization & ALS \footnote{The sweep-based DMRG optimization is alternating least squares
(ALS) algorithm in the TT context.\label{foot:als}}
& ALS \textsuperscript{\ref{foot:als}}
& N/A \footnote{The original ML-MCTDH method is a propagation method.} \\
4 & Integrator & N/A
\footnote{The original DMRG method is an eigen-value solver.\label{foot:solver}}
& N/A \textsuperscript{\ref{foot:solver}} 
& See References
\cite{wan03:1289,man08:164116,ven11:044135,lub04:355,lub15:917,klo17:174107,bon18:252,lin21:174108,lin21:174109,wan18:044119}  \\
5 & Interaction \footnote{It means interaction between fragments A and B of a complex system.} 
& short/long-range & short/long-range & short/long-range \\
6 & Application
\footnote{Only the original scenarios are given.}
& Various equations
\footnote{DMRG has capability to be applied to solve a wider range of equations
but not limited to solve the Schr{\"o}dinger equation.\label{foot:appli-dmrg}}
& Various equations \textsuperscript{\ref{foot:appli-dmrg}}
& Nuclear dynamics \\ 
\hline
\end{tabular}
\label{tab:dmrg-alg-ml-mctdh}
\end{sidewaystable}

%--tab.6------------Vibration----
\clearpage
\begin{sidewaystable}
\caption{%
A comparison of the approximations in resolving the vibration-structure
problem based on the vibrational self-consistent field (VSCF) method.
Resolving the vibration-structure problem, one could obtain ro-vibrational
eigen-states of the system. 
The first column gives various methods for resolving the vibration-structure
problem. The second column gives approximations in each method. The
third column gives the feature of size consistency. The last two column
gives counterpart in electron-structure theory as well as strategy for
resolving the equation of motion. We refer the reader to References
\cite{bow03:533,chr04:2140,pel10:20603,sei11:054119,fau15:134105,fau18:054104}
for details of VSCF and its variants.
}%
%\begin{ruledtabular}
\begin{tabular}{lllllllll}
\hline
Method \footnote{
Abbreviations ``VMP'' means vibrational M{\o}ller–Plesset perturbation
theory, ``PT2'' and ``VPT2'' second order perturbation theory, ``DC''
degeneracy corrected version,
``VCI'' vibrational configuration-interaction, ``ss'' state-specific version,
``VCC'' vibrational coupled-cluster, ``VMCSCF'' and ``VMRCC''
multi-configurational version of VSCF and VCC, respectively.}
&~~& \multicolumn{3}{c}{Vibration Structure} &~~& \multicolumn{3}{c}{Electron Structure}
 \\ \cline{3-5}\cline{7-9} 
       &~~& Approximation \footnote{
Abbreviations ``RSPT'' means Rayleigh-Schr{\"o}dinger perturbation theory.}
       &~~& Size &~~& Couterpart \footnote{
Abbreviations ``HF SCF'' means Hartree-Fock self-consistent field, ``MPPT''
M{\o}ller–Plesset perturbation theory, ``MP2'' second-order M{\o}ller–Plesset
perturbation theory, ``CI'' configuration interaction, ``CC'' coupled cluster,
``MCSCF'' multi-configurational self-consistent field, and ``MRCC''
multi-reference coupled cluster.
}
       &~~& Strategy   \\
        &&                 && Consistency && &&  \\
\hline
VSCF   && Hartree product of single-mode && no && HF SCF && variation    \\
VMP    && RSPT based on VSCF wave function &&no&& MPPT   && perturbation  \\
VSCF-PT2/VPT2&& second-order of VMP && no && MP2 && perturbation   \\
VSCF-DCPT2 && degeneracy corrected VPT2 && no && degenerated MP2 && degenerated RSPT   \\
HDCPT2 && hybrid version VSCF-DCPT2 && no && degenerated MP2 && degenerated RSPT  \\
VCI   && expansion using a set of VSCF wave functions && yes && full CI && variation  \\
ss-VCI&& state-specific configuration selected VCI && no && truncated CI && variation  \\
VCC&& coupled cluster expansion using VSCF wave functions && yes && full CC&&variation  \\
ss-VSCF && state-specific VSCF && no && --- && variation  \\
VMCSCF  && multi-configurational version of VSCF && no && MCSCF && variation \\
VMRCC   && multi-configurational version of VCC  && no && MRCC  && variation  \\
\hline
\end{tabular}
%\end{ruledtabular}
\label{tab:comp-vibrational-modes-vscf}
\end{sidewaystable}

%----tab.7
\clearpage
\begin{table}
\caption{%
Comparison and summary of quantum dynamics methods for various objects
in a molecular system, together with those for nucleon dynamcis ({\it
i.e.}, dynamcis of proton and neutron).
In these methods, a solution ansatz is substituted into the time-dependent
variational principle to obtain working equations. The solution ansatz
is expanded by products (sometimes called configurations) of one-dimensional
(or lower-dimensional) functions, called SPFs or MOs in different
scenarios. Correspondingly, the Hamiltonian operator is expanded into
a similar summation of products of one-dimensional (or lower-dimensional)
operators. Repeating such expansional processes, one can finally obtain
a hierarchical framework for various dynamics scenarios. The first column
gives the layer of the expansional form of solution ansatz, where the
SPFs or MOs in the deepest layer are further expanded by a given set
of basis functions. The second and third columns give methods for
wave-packet propagation for quantum molecular dynamics (QMD) and TD
electronic-structure (TDES) theory, respectively. We refer the reader
to Table \ref{tab:compare-nuclear-elec} for comparison of the QMD methods
with the TI electronic-structure methods which are essentially eigensolvers.
The fourth column gives methods for quantum nucleon dynamics (QND) which
are inherently within nuclear physics and hence beyond the scope of this
work. The rightmost column gives probable remarks. Here, we would like
to emphasize that the propagation methods for electron or nucleon must
preserve the permutation symmetry of TD wave function.
}%
\begin{tabular}{llllllllr}
\hline
Layer &~~& QMD &~~& TDES &~~& QND &~~& Remarks \\
\hline
zero && standard && TDHF \footnote{TDHF means time-dependent Hartree-Fock.\label{foot:tdhf}}
&& TDHF \textsuperscript{\ref{foot:tdhf}} && an impractical tool due to its \\
&&  &&    && && computational expense and \\
&&  &&    && && lack of correlation effects \\
one  && TDH && TDCI \footnote{TDCI means time-dependent configuration-interaction.}
%In practical implementations, The CI-type wave function is expanded in
%the CI basis, truncated at the level of either single (TDCIS) or single
%and double (TDCISD) excitations.}
&& && tool for improving time-dependent \\
&& && && && density functional theory (TDDFT) \\
&& && TDCC && TDCC && TDCC with $>2$ electronc does not \\
&& && && && converge to the full TDCI limit, and \\
&& && && && explains emergence of plasmon behavior \\
&& && TD-EOM-CC \footnote{TD-EOM-CC means TD equation-of-motion CC} &&
% which is often used with single and double excitations (namely TD-EOM-CCSD).} && 
&& a method for computing absorption spectrum  \\
two  && MCTDH && MCTDHF \footnote{MCTDHF employs the wave function in
the MCTDH form for dynamics of fermion particle.\label{foot:mctdhf}} 
&& MCTDHF \textsuperscript{\ref{foot:mctdhf}} && there exists MCTDHB for boson dynamics.  \\
multi && ML-MCTDH && TD-DMRG && TD-DMRG && MPS must preserve asymmetry  \\
\hline
\end{tabular}
\label{tab:comp-prop-methods}
\end{table}

%%%%%%%%%%%%%%%%%%%%%%%%%%%%%%%%%%%%%%%%%%
% FIGURES
%%%%%%%%%%%%%%%%%%%%%%%%%%%%%%%%%%%%%%%%%%%
\clearpage
\section*{Figure Caption}

\figcaption{fig:framework}{
Systematic illustration of the quantum dynamics methods. The black
words represent computational goals or results, while the red words
represent computational techniques. The arrows represent the direction
from the input to the computational results. Designing a set of
coordinates, the kenetic energy operator (KEO) can be derived through
the polyspherical approach. Then, the database is constructed through
{\it ab initio} energy calculations on a set of sampled geometries.
The potential energy surface (PES) is then constructed in an appropriate
form. With the Hamiltonian operator, extensive wavepacket propagations
can be launched on the basis of the rovibrational eigenstate of the
reactant which can be computed by relaxation of a guess funciton. Then,
the initial wave function is computed by assigning a given value of
momentum to the initial eigenstate. Having time-dependent wave function,
flux analysis and expectation calculations are launched to compute
reaction probability and time-dependent expectations. The present work
presents a computational framework for wavepacket propagation employing
the multiconfiguration time-dependent Hartree (MCTDH) and its multilayer
extension (ML-MCTDH) methods, where both the Hamiltonian and wave function
are expanded in hierarchical form. The blue arrows belong to construction
of the Hamiltonian as given in Sections \ref{sec:coord-keo} and \ref{sec:pes-cpd}. 
The black arrows belong to the wavepacket propagation and analysis of
the wave function as give in Section \ref{sec:working}.
}

\figcaption{fig:hier-coord}{
Hierarchical scheme to design coordinates, where a circle indicates a
frame and a box represents the coordinate in the deepest layer. The first
layer separates the E$_2$ frame from relative coordinates with respect
to the SF frame, denoted by E$_2$/SF. Dividing the system into several
subsystems, a BF frame is designed for each subsystem. The second layer
of the E$_2$ frame separates all of BF frames from each other. Moreover,
relative coordinates of BF frame with respect to the E$_2$ frame (denoted
by BF/E$_2$) is separated from other BF frames in the second layer.
Repeating this separating processes, we can obtain all of coordinate
frames. The deepest layer gives (1) either Jacobi or Radau coordinates
to describe motions of fragment and (2) Euler angles or relative
positions to describe relative motions.
}

\begin{figure}
 \centering
  \includegraphics[width=24cm,angle=90]{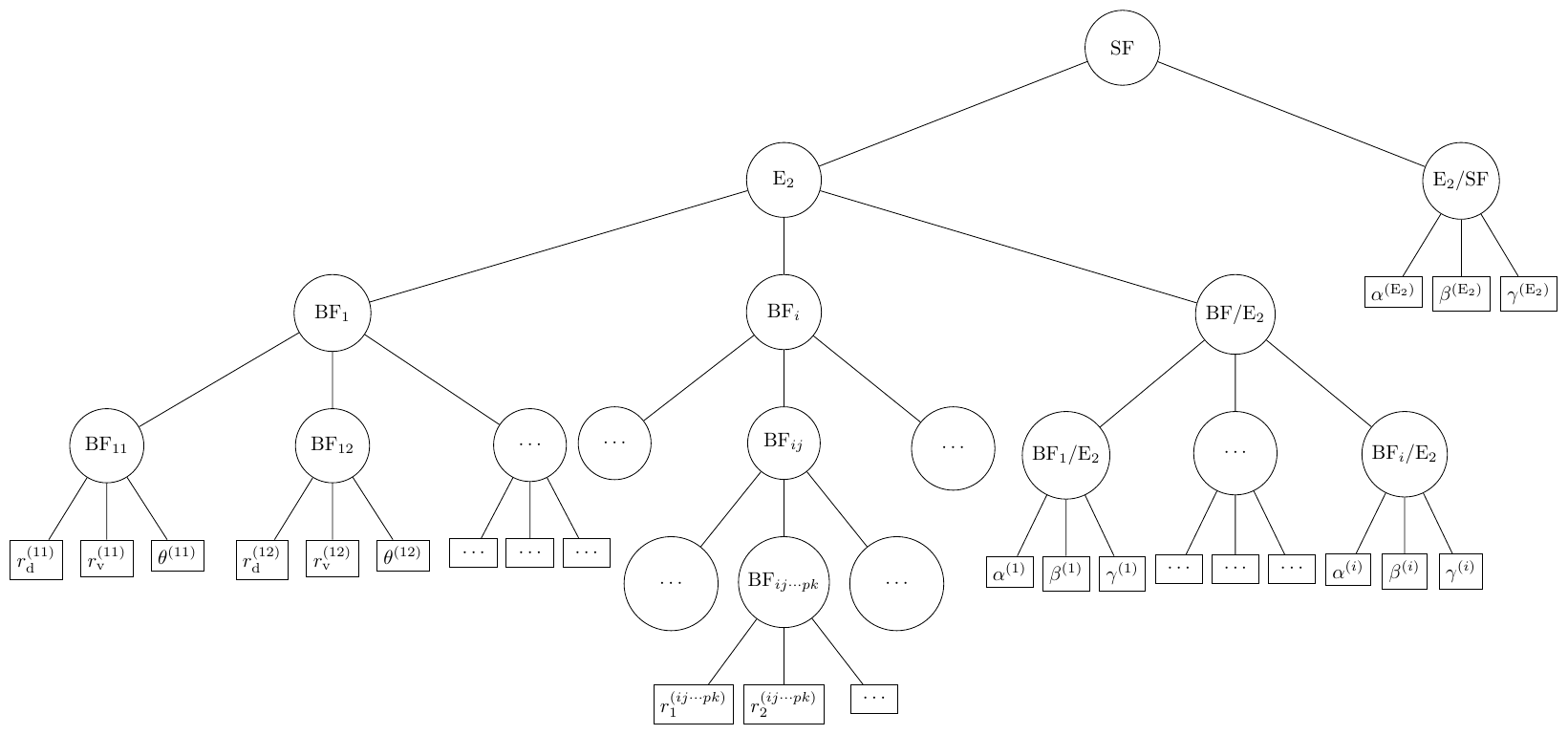}
   \caption{\figfoot}
    \label{fig:hier-coord}
     \end{figure}

\figcaption{fig:ml-expansion-examples}{
Hierarchical coordinates frames of (a) CO/Cu(100), (b)
H$_2$O/Cu(111), and (c) HO$_x$, which are designed according to Figure
\ref{fig:hier-coord}. For CO/Cu(100) and H$_2$O/Cu(111), surface
scattering processes are described by E$_2$ frame which is separated
into (1) BF$_{\mathrm{CO}}$ or BF$_{\mathrm{wat}}$ for molecular
motions, (2) relative frame BF$_{\mathrm{CO}}$/E$_2$ or
BF$_{\mathrm{wat}}$/E$_2$ for relative positions, and (3)
BF$_{\mathrm{Cu}}$ for surface atoms which is divided into BF$_1$
and $\{\mathrm{BF}_i\}$ for top and non-top atoms, respectively. The
relative frame E$_2$/SF is further required to describe molecular
motions with respect to surface. For HO$_x$, if the total angular
momentum of the whole system is set to be zero, the Euler angles in
the relative frame E$_2$/SF are all equal to zero which implies that
the E$_2$ and SF frames are identical. The OH and HO$_2$ fragments
are described by BF$_{\mathrm{OH}}$ and BF$_{\mathrm{HO}_2}$,
respectively, together with relative frames BF$_{\mathrm{OH}}$/E$_2$
and BF$_{\mathrm{HO}_2}$/E$_2$. The relative frame
BF$_{\mathrm{OH}}$/BF$_{\mathrm{HO}_2}$ is required to define
principal component of reaction coordinate.
}

\begin{figure}
  \centering
   \includegraphics[width=16.5cm]{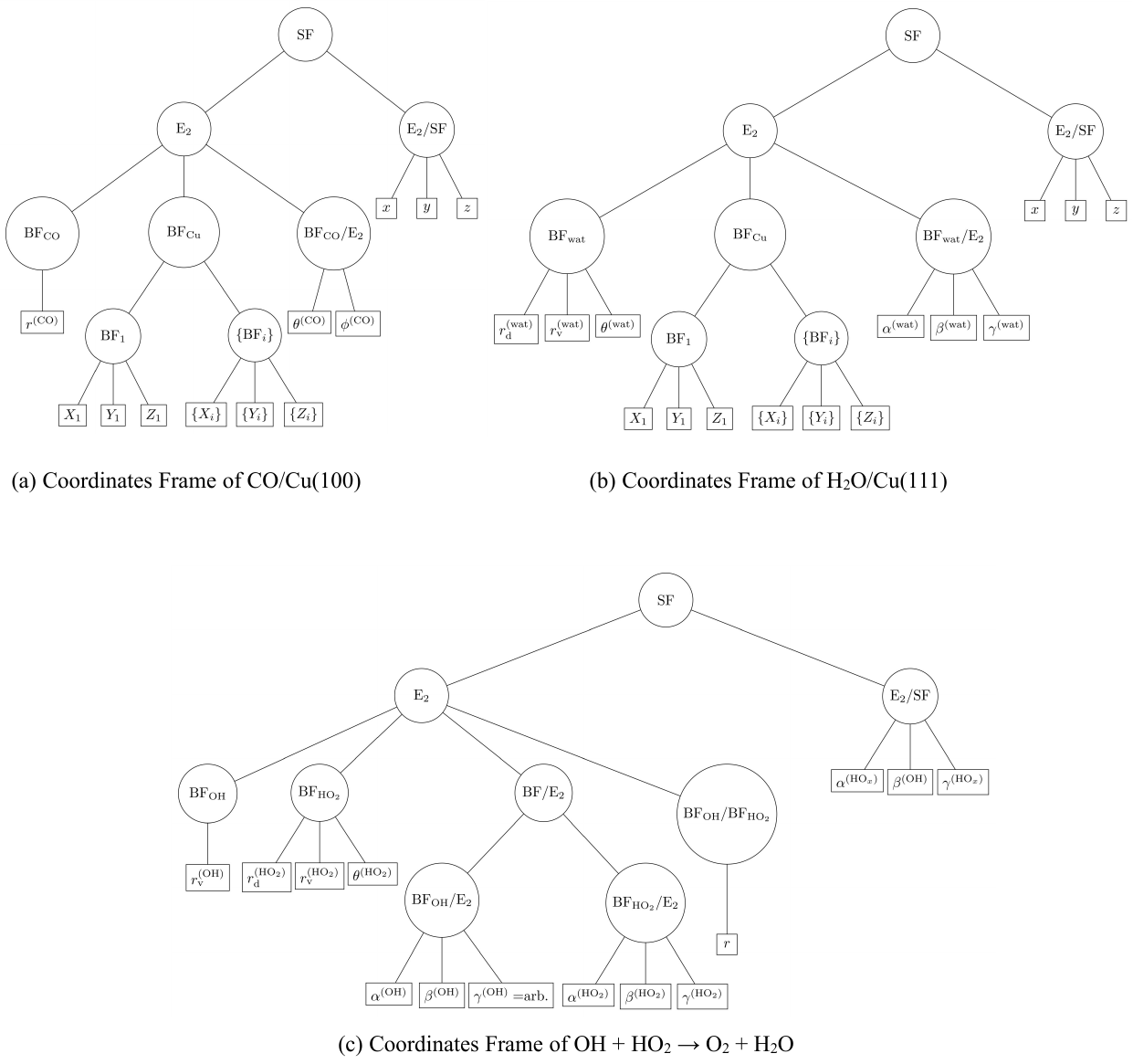}
    \caption{\figfoot}
     \label{fig:ml-expansion-examples}
	    \end{figure}

\figcaption{fig:wf-ml-mctdh}{
ML-MCTDH wave function structures (ML-tree structures) of the (a)
CO/Cu(100) \cite{men21:2702}, (b) H$_2$O/Cu(111) \cite{son22:6047},
and (c) HO$_x$ \cite{son24:597} systems. These ML-tree structures are
iteratively optimized to indicate fast and well convergence of
ML-MCTDH calculations \cite{men12:134302,men13:014313}. Numbers of
SPFs and primitive basis functions are given next to lines. Since the
total angular momentum of the HO$_x$ system is set to be zero
\cite{son24:597}, its dynamical model unnecessarily incorporates the
Euler angles $\{\alpha^{(\mathrm{HO}_x)},\beta^{(\mathrm{HO}_x)},\gamma^{(\mathrm{HO}_x)}\}$
of the relative frame E$_2$/SF shown in Figure \ref{fig:ml-expansion-examples}(c).
}

\figcaption{fig:results-flux-exp}{
ML-MCTDH results of (a) and (b) the CO/Cu(100) system, (c) the
H$_2$O/Cu(111) system, and (d) the HO$_x$ system. Subfigure (a)
illustrates time-dependent position expectations along out-of-plane
coordinates of CO (the solid lines) and top copper atom (the dashed
line), while subfigure (b) shows sticking probabilities. The red and
blue lines represent ML-MCTDH results from rigid (6D) and nonrigid
(21D) surface models, respectively. Reprinted with permission from
Reference \cite{men21:2702}. Copyright 2021 American Chemical Society.
Subfigure (c) illustrates dissociative sticking probabilities computed
for rigid surface model. The colored lines were explained in Reference
\cite{son22:6047}. Reprinted with permission from Reference \cite{son22:6047}.
Copyright 2022 American Chemical Society. Subfigure (d) illustrates
reactive probabilities for the OH + HO$_2$ $\to$ O$_2$ + H$_2$O reaction.
Reprinted with permission from Reference \cite{son24:597}. Copyright
2024 American Chemical Society.
}

\figcaption{fig:h-gra-dyna}{
The 2D reduced density of the wave function for the $z$ directions of
the H atom and the $q_1$ normal mode at $10$, $38$, $46$, $58$, $78$,
$82$, $120$, and $260$ fs during simulation on the PES 2D cut from the
DOFs correspondent for the simulation of the H atom with an initial
kinetic energy of $0.96$ eV. We refer the reader to References \cite{shi23:194102,shi25:1896}
for numerical details and explaination. Reprinted with permission from
Reference \cite{shi23:194102}. Copyright 2023 American Institute of
Physics.
}

\figcaption{fig:classif}{
Diagrammatic sketch for classification of the methods derived from
time-independent (denoted by ``TI'' in the left panel) or time-dependent
(denoted by ``TD'' in the right panel) variational principle and
organized hierarchically with function (upper panel) or tensor (lower
panel) representation. The black words indicate methods for molecular
dynamics, while the red words indicate electronic-structure techniques.
These methods exhibit mathematical tree-structures
whether expressed in grid-based function representation or second-quantized
representation or tensor representation, demonstrating deep relations
among these methods. We would like to emphasize that the the classification
by ``function'' and ``tensor'' is informal serving only as illustration.
For instance, the original way of DMRG views the wave function in terms
of renormalized functions and hence should be moved to the upper panel.
This comparative framework reveals
how TI or TD variational principle unifies different representations,
suggesting potential cross-methodological applications, in particular
in addressing systems with correlated DOFs. This implies that insights
from tensor network (TN) or tree tensor network (TTN) could inform
those indicated by wave function,
and vice versa, highlighting opportunities for methodological cross-fertilization
between electronic-structure and quantum molecular dynamics domains.
}

\figcaption{fig:ring-polymer-ueff}{
Diagrammatic sketch of the ring-polymer potential given by Equation
\eqref{eq:path-integral-for-004-pimd}. The blue symbols represent $P$
virtual systems which are all identical and share the same Hamiltonian
operator. The gray line represents the potential function $V(x)$, where
the abscissa and ordinate axes represent coordinate $x$ and potential
$V$, respectively. The red lines represent harmonic interaction between
adjacent virtual systems while $P$ is number of virtual systems. The
virtual systems can be seen as those assumed in the path integral. These
interacted virtual systems form an ensemble if $P\to\infty$.
}

%--fig.1------------------------------------
\clearpage

%--fig.2---------------------
\clearpage

%--fig.3----------------
\clearpage

%%%fig.4%%%%%%%%%%%%%%%%%%%
\clearpage
 \begin{figure}
  \centering
   \includegraphics[width=19.5cm,angle=90]{./fig4-ml-tree}
    \caption{\figfoot}
     \label{fig:wf-ml-mctdh}
     \end{figure}

%---fig.5------------------
\clearpage

%---fig.6-----------------------
\clearpage

%--fig.7---------
\clearpage

%-fig.8-----------ring-polymer----------
\clearpage

%-------------------------------------------------
% References
%--------------------------------------------------
\clearpage
%\bibliographystyle{../bibtex/mybibu}
%\bibliography{../bibtex/refs}

\end{document}